\documentclass[11pt,a4paper]{article}

\usepackage{microtype}
\usepackage{amsmath, amsfonts,amssymb,mathtools,nicefrac,nccmath,cases,physics}
\usepackage{bm}
\usepackage{soul}
\usepackage{dsfont}
\usepackage{booktabs,multirow}
\usepackage{tabularx}
\usepackage[usenames,dvipsnames,table]{xcolor}
\usepackage{colortbl}
\usepackage{caption}
\usepackage{subcaption}
\usepackage{float}
\usepackage{hhline}
\usepackage{enumitem}
\usepackage{cite}
\numberwithin{equation}{section}
\numberwithin{table}{section}
\usepackage[linktocpage=true]{hyperref}
\hypersetup{colorlinks=true,linkcolor=colorloc3,citecolor=colorloc3,urlcolor=colorloc3}
\usepackage{afterpage}

\def\hybrid{\topmargin -20pt    \oddsidemargin 0pt
	\headheight 0pt \headsep 0pt
	\textwidth 6.5in        
	\textheight 9in         
	\textwidth 6.25in       
	\textheight 9 in       
	\marginparwidth .875in
	\parskip 5pt plus 1pt 
	\jot = 1.5ex
}
\hybrid

\usepackage[framemethod=default]{mdframed}

\newmdenv[skipabove=10pt,
skipbelow=7pt,
rightline=false,
leftline=true,
topline=false,
bottomline=false,
linecolor=colorloc2,
backgroundcolor=colorloc1!5,
innerleftmargin=4pt,
innerrightmargin=0pt,
innertopmargin=0pt,
leftmargin=2pt,
rightmargin=0pt,
linewidth=2pt,
innerbottommargin=0pt,
frametitlebackgroundcolor=colorloc2]{lbBox}

\usepackage{CormorantGaramond}

\definecolor{colorloc1}{RGB}{164,42,46} 
\definecolor{colorloc2}{RGB}{100,100,100} 
\definecolor{colorloc3}{RGB}{204,119,34}  
\definecolor{colorloc4}{RGB}{25,25,112}  
\definecolor{colorloc5}{RGB}{0,128,128}  

\usepackage{sectsty}
\usepackage{titlesec}

\sectionfont{\color{colorloc4}\Large} 

\subsectionfont{\color{colorloc2}\large} 

\subsubsectionfont{\color{colorloc5}} 

\usepackage[linktocpage=true]{hyperref}
\hypersetup{colorlinks=true,linkcolor=colorloc5,citecolor=colorloc4,urlcolor=colorloc4}

\newcommand{\im}{\mathbf{i}}

\begin{document}


  \baselineskip=14pt
  \parskip 5pt plus 1pt

	\vspace*{-1.5cm}
  
  \vspace{4cm}
  \begin{center}        

    {\color{colorloc4}  \huge Machine learning the breakdown \\ of tame effective theories}   
  \end{center}
  
  \vspace{0.2cm}
  \begin{center}        
    {\large  Stefano Lanza}
  \end{center}
  
  \begin{center}  
    \emph{II. Institut f\"ur Theoretische Physik, Universit\"at Hamburg,\\
    	Luruper Chaussee 149, 22607 Hamburg, Germany}
    \\\vspace{0.2cm} \emph{\&} \\\vspace{0.2cm}
    \emph{Institute for Theoretical Physics, Utrecht University,\\
    	Princetonplein 5, 3584 CC Utrecht, The Netherlands}
  \end{center}
  
  \vspace{2.5cm}
  
  
\begin{abstract}
	Effective field theories endowed with a nontrivial moduli space may be broken down by several, distinct effects as the energy scales that are probed increase. 
	These may include the appearance of a finite number of new states, or the emergence of an infinite tower of states, as predicted by the Distance Conjecture. 
	Consequently, the moduli space can be partitioned according to which kind of state first breaks down the effective description, and the effective-theory cutoff has to be regarded as a function of the moduli that may abruptly vary in form across the components of the partition.
	In this work we characterize such a slicing of the moduli space, induced by the diverse breakdown mechanisms, in a two-fold way.
	Firstly, employing the recently formulated Tameness Conjecture, we show that the partition of the moduli space so constructed is composed only of a finite number of distinct components.
	Secondly, we illustrate how this partition can be concretely constructed by means of supervised machine learning techniques, with minimal bottom-up information.
\end{abstract}

\thispagestyle{empty}
\clearpage
  
\setcounter{page}{1}
  
  
\newpage

\tableofcontents

\newpage

\section{Introduction}
\label{sec:Introduction}

Effective field theories have been serving as powerful tools for physicists, allowing for focusing solely on the interactions occurring at the energy scales of interest, while neglecting other phenomena that take place in different regimes.
As such, the definition of an effective theory is intimately related to the energy cutoff, being the energy scale at which the effective description is rendered invalid due to the emergence of new interactions, possibly mediated by new states.
In turn, the knowledge of the effective theory cutoff -- namely at which energy scale it is identified and by which new interactions it is determined -- carries pivotal information about how the effective theory may be completed in the ultraviolet regime.

A prime example is Fermi's four-fermion theory, which describes weak interactions among fermions for sufficiently small energy scales.
If the energy scales that are probed increase, reaching energies as high as $\Lambda \simeq 80\, \text{GeV}$, the theory is not able to capture the several interactions occurring at those scales, with the reason being the emergence of the non-negligible $W$ and $Z$-bosons interactions, which the theory ought to account for.
Indeed, we do know that Fermi's theory can be completed within the Standard Model, for which the former serves as an effective theory valid below its cutoff, that is indeed set by the mass of the $W$-bosons $\Lambda_{\text{\tiny EFT}} = m_W \simeq 80\, \text{GeV}$.

In effective theories that are endowed with a moduli space, the identification of a cutoff is more subtle and less clear.
Indeed, the vacuum expectation values of the moduli determine the strength of the couplings that govern the interactions and, along with those, the masses of the states that, descending from the ultraviolet completion, could break down the description.
Consequently, in general, effective theories with a moduli space are not characterized by a single, univocally identified cutoff: rather, in different regions of the moduli space the effective description might be broken down by \emph{different kinds of states}, and at \emph{different energy scales}.

Effective field theories originating from string theory are naturally endowed with moduli spaces with rich geometrical structures, and the definition of their cutoff is tied to the geometry of the moduli space.
In this regards, towards the boundaries of the moduli space, some `hard' effective-theory breakdowns may occur, such as the one predicted by the Distance Conjecture \cite{Ooguri:2006in}.
The conjecture predicts that in any effective theory that admits a completion within string theory an infinite tower of states becomes light as boundaries of the moduli space located at infinite distance are approached.
Thus, towards such infinite-distance boundaries, the effective-theory cutoff should be consistently reduced, in order to keep the infinite tower integrated out, thus getting lower and lower as said boundary is reached.

The Distance Conjecture, as a source of effective description breakdown, is thus invariably affected by the aforementioned ambiguities regarding the definition of an effective-theory cutoff.
Indeed, in different regions of the moduli space different infinite towers may first break down the effective theory, for they are the lightest ones in those regions, and the mass of their lightest state may serve as cutoff.
This phenomenon is common to several effective theories stemming from string theory -- see \cite{Hebecker:2017lxm,Grimm:2018ohb,Grimm:2018cpv,Gendler:2020dfp,Palti:2021ubp, Lee:2018urn,Lee:2018spm,Lee:2019tst,Lee:2019xtm,Lee:2019wij,Klaewer:2020lfg, Lanza:2020qmt,Lanza:2021udy,Grimm:2022sbl,Etheredge:2023odp} for a sample of works.
In general, however, it is not clear how many towers are needed in order to realize the Distance Conjecture throughout the moduli space, and how to systematically partition the moduli space according to the regions where each of these towers is relevant.

These ambiguities in the definition of an effective-theory cutoff can be encoded within the following key, general questions:
\begin{itemize}
	\item given a moduli space, how \emph{many} different states, stemming from the ultraviolet completion, could lead to the breakdown of the effective theory? Are they \emph{infinite} in number, or do they belong to a \emph{finite} set?
	\item how can we \emph{concretely} determine how the effective theory is broken down throughout the moduli space, and what is the \emph{minimal} information required to achieve this? 
\end{itemize}
Addressing these questions is the core of this work, and the strategy that we will employ to tackle them is two-fold, relying on both a top-down and a bottom-up viewpoint.

The top-down approach that we will pursue is motivated by the recently formulated Tameness Conjecture \cite{Grimm:2021vpn}.
The conjecture states that the couplings that appear in any consistent effective field theory have to be \emph{tame}, and belong to specific families of functions -- these concepts will be reviewed in Section~\ref{sec:Tame_DC_Tameness}.
Exploiting the conjecture, in \cite{Grimm:2022sbl} it was shown that the realization of the Distance Conjecture requires only a \emph{finite} number of different infinite towers of states in any tame effective theory.
In this work, we will generalize these results, and illustrate that in effective theories governed by tame couplings, only a \emph{finite} number of different types of states is sufficient to define the cutoff throughout the moduli space.

The bottom-up viewpoint that we will deliver offers a concrete method to compute the slicing of the moduli space according to how the effective theory cutoff is defined.
Achieving this requires assuming some additional, though minimal information about the effective theory: we need to know how the effective theory is broken down at \emph{some} points of the moduli space.
Then, we feed this information, serving as \emph{dataset}, to some supervised machine learning algorithms - specifically, we will employ the $k$-nearest neighbor algorithm, and a linear support vector machine algorithm.
The algorithms, utilizing the above dataset, are able to reconstruct the regions of the moduli space where a given state is the one that determines the effective-theory cutoff.

The two viewpoints are not independent of one another, rather they are complementary.
In fact, the application of the machine learning algorithms to the problems at hand is rooted in the fact that the couplings of the effective theory are tame, and only a finite number of states contribute to the definition of the effective-theory cutoff. 
If the number of different kinds of states defining the cutoff were infinite, the application of any machine learning algorithms would be pointless and unsuccessful, for one would incur the infamous `\emph{halting problem}'.
The complementarity of the two pictures will be additionally illustrated in Section~\ref{sec:Ex_Tor} with a concrete example, namely the computation of the cutoff-induced slicing of the moduli space of a four-dimensional effective theory obtained after compactifying Type IIB string theory over a toroidal orbifold.

This work is articulated as follows. 
In Section~\ref{sec:Tame_DC} we overview the statements of the Tameness Conjecture, and its implications for effective field theory in general; we also recall how the conjecture may be helpful in investigating how effective theories are broken down due to the emergence of infinite towers of states, reviewing the main findings of~\cite{Grimm:2022sbl}.
Section~\ref{sec:ML_and_DC} collects the foundational ideas of this work, which are presented from two complementary perspectives: we illustrate how the tameness of the effective theory implies that the moduli space is partitioned in a finite number of subsets, according to the kind of states that determines the cutoff there; then, we show how such a partition can be obtained purely from a bottom-up perspective via two supervised machine learning techniques, the $k$-nearest neighbor algorithm and the linear support vector machine algorithm.
In Section~\ref{sec:Ex_Tor} we show how the ideas presented in Section~\ref{sec:ML_and_DC} concretely apply to a four-dimensional effective theory obtained after compactifying Type IIB string theory over a toroidal orbifold.
Finally, in Appendices~\ref{sec:k-near_review} and~\ref{sec:LSMVMreview} we briefly overview the $k$-nearest neighbor algorithm and the linear support vector machine algorithm employed in Section~\ref{sec:ML_and_DC}.

\section{The Distance Conjecture in Tame Effective Field Theories}
\label{sec:Tame_DC}

The Distance Conjecture \cite{Ooguri:2006in} is one of the founding statements from which the Swampland Program sparkled.
The conjecture aims at identifying generic features that any consistent effective field theory should be equipped with towards specific field space boundaries.
Although such genericity makes the conjecture appealing, it also renders the conjecture hard to prove, even in simple, concrete frameworks.

The claims of the Distance Conjecture can however be better addressed if one invokes another recently formulated conjecture, the Tameness Conjecture \cite{Grimm:2021vpn}.
The latter, constraining the functional form of the couplings that appear in any consistent effective field theory, not only can help uncover novel features that effective field theories exhibit near the field space boundaries; it also delivers recipes that simplify tests of the Distance Conjecture \cite{Grimm:2022sbl}.

In this section, in order to set the ground for the forthcoming discussion, we briefly recall the core statements of the Distance Conjecture, and the limitations that emerge when trying to test it. 
Then, we will recall how the Tameness Conjecture can help better address the Distance Conjecture by overviewing the main results of \cite{Grimm:2022sbl}.

\subsection{The Distance Conjecture, and its testing limitations}
\label{sec:Tame_DC_Limit}

For the sake of generality, we consider an effective theory formulated in $D$ spacetime dimensions.
The assumptions on the effective field theory are minimal: beside assuming that the theory is coupled to gravity, we solely assume the theory to be endowed with a non-trivial $N$-dimensional, classical moduli space $\mathcal{M}$.
A local patch within $\mathcal{M}$ can be parametrized by the coordinates $\varphi^A$, with $A = 1, \ldots, N$, the moduli of the theory.
Since we shall focus solely on single patches of the moduli space, in which the effective theory is well-defined, with an abuse of nomenclature, we will oftentimes confuse a point in $\mathcal{M}$ with its coordinate representation as $\varphi^A$. Additionally, within the given patch, the moduli space is endowed with the local metric $G_{AB}(\varphi)$.

The effective action describing the dynamics of the moduli $\varphi^A$, defined for a finite cutoff $\Lambda_{\text{\tiny EFT}}$ include the following contributions:
\begin{equation}
	\label{TameDC_Sgen}
	S^{(D)}_{\text{\tiny EFT}} = M_{\text{P}}^{D-2} \int {\rm d}^Dx\, \sqrt{-g} \left( \frac{1}{2} R - \frac12 G_{AB}(\varphi) \partial_\mu \varphi^A \partial^\mu \varphi^B + \ldots \right)\, .
\end{equation}
Here, $M_{\text{P}}$ is the $D$-dimensional Planck mass, $g = \det g_{\mu\nu}$, with $g_{\mu\nu}$ being the spacetime metric, and $R$ is the spacetime Ricci scalar.
In \eqref{TameDC_Sgen}, the dots denote some additional contributions that may include other fields, and are not relevant for the discussion that unfolds below;
however, importantly, in this work we assume that the fields $\varphi^A$ are not constrained by any scalar potential.

The Distance Conjecture, first formulated in \cite{Ooguri:2006in}, delivers key statements about the geometry of the moduli space $\mathcal{M}$ for any effective field theory of the form \eqref{TameDC_Sgen} that stems from a proper quantum gravity theory.
The content of the conjecture can be divided into two parts:

	\noindent\textbf{Non-compactness of $\mathcal{M}$:} for any $C > 0$ and for any point in $\mathcal{M}$ parametrized by the coordinates $\varphi^A_0$, there exists a point, parametrized by the coordinates $\varphi^A$, such that
	\begin{equation}
		\label{TameDC_d>C}
		d(\varphi_0 , \varphi)  > C\,,
	\end{equation}
	where $d(\varphi_0 , \varphi)$ is the length of a geodesic path stretching between the point $\varphi^A_0$ and the point $\varphi^A$.
	Namely, parametrizing such a geodesic path as $\varphi^A(\sigma)$, where $\sigma \in [0,1]$, and with the endpoints $\varphi^A(0) = \varphi^A_0$ and $\varphi^A(1) = \varphi^A$, we define
	\begin{equation}
		\label{TameDC_d_def}
		d(\varphi_0, \varphi) = \int_0^1 \sqrt{G_{AB} (\varphi) \frac{{\rm d} \varphi^A}{{\rm d} \sigma} \frac{{\rm d} \varphi^B}{{\rm d} \sigma}}\, {	\rm d} \sigma\,.
	\end{equation}
	Said differently, there exist some \emph{boundary} points in $\mathcal{M}$, that we label as $\varphi^A_{\rm b}$, that are located at infinite geodesic distance from any other point in $\mathcal{M}$.
	
	\noindent\textbf{Emergence of an infinite tower of states:} there exists an infinite tower of states that become exponentially light as a boundary located at infinite field distance is reached.
	Specifically, let us call $M_n(\varphi)$ the moduli-dependent masses of the states constituting the tower, with $n \in \mathbb{N}$ the index labeling the states in the tower.
	Then, the Distance Conjecture asserts that the masses of the tower located at two points $\varphi^A$ and $\varphi_0^A$ are related as
	\begin{equation}
		\label{TameDC_Mn}
		M_n(\varphi) \sim M_n(\varphi_0)\, e^{ -\lambda d(\varphi_0, \varphi)}\,,
	\end{equation}
	with $\lambda$ a real parameter. 
	
	The emergence of such an infinite tower of states \emph{badly} breaks down the effective description that we started with: indeed, assuming the effective theory to be characterized by a finite cutoff $\Lambda_{\text{\tiny EFT}}$ for any point $\varphi^A$, an infinite number of states should ideally be included close enough to the infinite-distance boundary.
	Moreover, it is worth stressing that, allegedly, the infinite tower with states behaving as in \eqref{TameDC_Mn} is the \emph{relevant} one that breaks down the effective field theory: there may be other infinite tower of states that also break down the effective field theory, but they are more massive than said relevant one.

In order to corroborate the conjecture, some tests have been performed in the literature during the last years.
However, as remarked earlier, excluding some trivial, one-dimensional cases, tests have been revealed to be hard, even in simple setups with just a couple of moduli reaching the boundary.
In fact, while the non-compactness of the moduli space seems a property that is commonly shared by all the moduli spaces that are obtained by compactifying string theory over some internal manifold, proving the emergence of an infinite tower of states towards an infinite-distance boundary, exhibiting the behavior \eqref{TameDC_Mn}, is far from trivial.

In principle, tests of the latter property should proceed in two steps: first, one should catalog all the the possible `candidate' microscopic states that could deliver an infinite tower emerging in the effective field theory; then, among these candidates, one should find the relevant one that breaks down the effective theory before the others. 
Secondly, one should show that the behavior of the masses of the states within the relevant candidate tower behave as in \eqref{TameDC_Mn} for some parameter $\lambda$.
It is then clear that, in order to perform such checks, not only a detailed knowledge of the microscopic theory is required; additionally, explicit checks of the behavior \eqref{TameDC_Mn} require an explicit computation of the geodesic distance \eqref{TameDC_d_def} which might not be analytically doable.

Furthermore, it is worth remarking that, in its original formulation, the Distance Conjecture did not cover several issues that may be pivotal in addressing concrete checks thereof.
For instance, the emergence of an infinite tower of states should occur towards an infinite distance boundary \emph{regardless} of how the boundary is approached.
Namely, one can approach the boundary by following any arbitrary path, and may find a tower of states becoming massless with a fall-off of the masses still regulated by \eqref{TameDC_Mn}; however, choosing another path, another, different tower of states may be the relevant one for realizing the Distance Conjecture, being lighter than the former.
Indeed, in this regard, the Distance Conjecture does not specify that the tower that is relevant along any path leading to a given boundary has to be unique.
Actually, since one can approach a boundary along infinite paths, and each may exhibit a different, relevant tower, the towers needed for realizing the conjecture may be infinite in number!

\subsection{The Tameness Conjecture}
\label{sec:Tame_DC_Tameness}

The general, aforementioned issues can be addressed employing the Tameness Conjecture  \cite{Grimm:2021vpn}, whose statements we briefly overview. 
The Tameness Conjecture states the following:

\noindent\textbf{Tameness Conjecture:} All the effective theories valid below a fixed finite energy cut-off scale can be labeled by a definable parameter space and are endowed with scalar field spaces and coupling functions that are definable in an o-minimal structure.

In this work we will not attempt to deliver a mathematical review of o-minimal structures, for which we refer to the original Tameness Conjecture paper \cite{Grimm:2021vpn} or the follow-up work \cite{Grimm:2022sbl}.  
Rather, we outline the consequences for the effective field theories that come by assuming the Tameness Conjecture.

For the sake of concreteness, consider a moduli-dependent coupling $y$ appearing in the effective field theory, and assume that the o-minimal structure of interest is $\mathbb{R}_{\text{an,exp}}$\footnote{It can be shown that in several effective field theories originating from string theory all the couplings entering the theory belong to this structure. In particular, this is the case of the couplings of the vector multiplet sector of the four-dimensional $\mathcal{N}=2$ F-theory/Type IIB effective theories \cite{Grimm:2021vpn,Grimm:2022sbl}. However, string theory may deliver more general o-minimal structures \cite{Grimm:2023xqy}.}.
For instance, in the effective theories of the form \eqref{TameDC_Sgen} the coupling $y$ may be one of the elements of the field space metric $G_{AB}(\varphi)$, or one of the masses of the states within the tower \eqref{TameDC_Mn} realizing the Distance Conjecture.
Then, if the coupling $y$ resides in a tame effective theory, supported by an $\mathbb{R}_{\text{an,exp}}$ o-minimal structure, the functional form of $y$ in terms of the moduli can be obtained by solving the following set of equations and inequalities:
\begin{equation}
	\label{TameDC_Loci}
	\begin{aligned}
		\exists\; x_1, \ldots, x_l: \qquad &P_I(\varphi,\lambda,x,y, f_1, \ldots, f_m, e^\varphi, e^\lambda, e^x, e^y)=0\,,
		\\
		&Q_J(\varphi,\lambda,x,y, f_1, \ldots, f_m, e^\varphi, e^\lambda, e^x, e^y)>0\,,
	\end{aligned}
\end{equation}
In the locus above, $P_I$ and $Q_J$ are generic polynomials of their arguments.
The latter include: the moduli $\varphi^A$ appearing in the effective field theory; some external parameters $\lambda^\kappa$ (which may be, for instance, internal geometrical parameters related to the specific compactification); the coupling $y$; some auxiliary variables $x_1, \ldots, x_l$, which might be needed solely to recast the coupling $y$ as the locus \eqref{TameDC_Loci}; some restricted analytic functions\footnote{We refer to \cite{Grimm:2022sbl} for a definition of restricted analytic functions, with examples thereof.} $f_1, \ldots, f_m$, with all the aforementioned variables as arguments.

As remarked in \cite{Grimm:2022sbl}, although the loci \eqref{TameDC_Loci} deliver general couplings of tame effective theories, in concrete string theory settings it is enough to focus on specific, though large families of tame couplings: namely, the `\emph{polynomially tamed couplings}' or its subset of `\emph{monomially tamed couplings}'.

In order to define these families of couplings, it is convenient to be more specific about the patch of the moduli space where our effective field theory is defined. 
We are interested in \emph{near-boundary} patches of the moduli space, that cover the neighborhood of a boundary of the moduli space.
Within a patch so defined, which we call $\mathcal{E}$, we separate the moduli fields $\varphi^A$ into two distinct sets:
\begin{description}
	\item[Saxions $s^i$] ($i = 1, \ldots, n$): these fields span \emph{non-compact} domains; we shall assume that $s^i > 1$, and that the field space boundary around which our effective theory is defined is reached as $s^i \to \infty$ $\forall\, i$;
	\item[Axions $a^\alpha$] ($\alpha = 1, \ldots, N - n$): these are fields that span \emph{compact} domains; after properly normalizing them, we will assume that $a^\alpha \in [0,1[$, with the identification $a^\alpha \simeq a^\alpha + 1$.
\end{description}

In sum, the patch $\mathcal{E}$ acquires the following form:
\begin{equation} 
	\label{TameDC_cE}
	\mathcal{E} = \{ |a^\alpha| < 1,\ \  s^1 ,  s^2 , \ldots , s^n > 1 \}\, .
\end{equation}
We consider a positive definite coupling $y$ defined over the patch $\mathcal{E}$; we then identify:

	\noindent\textbf{Monomially tamed couplings:} these are couplings that admit the following , general expansion in terms of the moduli
	\begin{equation}
		\label{TameDC_y_exp}
		y(s,a) = \sum_{ \mathbf{m}}\rho_{\mathbf{m}}(e^{-s^i},a^\alpha)   (s^1)^{m_1} \cdots (s^n)^{m_n}\, , 
	\end{equation}
	where $\rho_{\mathbf{m}}(e^{-s^i},a^\alpha)$ are restricted analytic functions of their arguments, and $\mathbf{m} = (m_1, \ldots, m_n) \in \mathbb{Z}^n$, and they are such that they are well approximated by a monomial in $\mathcal{E}$.
	The latter property is concretely expressed by requiring that
	\begin{equation}
		\label{TameDC_mon_tame}
		y(s,a) \sim (s^1)^{k_1} \cdots (s^n)^{k_n} \qquad \text{on} \quad \mathcal{E}\ , 
	\end{equation}
	for some $k_1, \ldots, k_n \in \mathbb{Z}$.
	The symbol `$\sim$' has to be understood as a double-boundness statement as follows: there exist two real, positive parameters $C_1$, $C_2$ such that
	$ C_1 (s^1)^{k_1} \cdots (s^n)^{k_n} < y < C_2 (s^1)^{k_1} \cdots (s^n)^{k_n}$ in $\mathcal{E}$.
	
	\noindent\textbf{Polinomially tamed couplings:} analogously to monomially tamed couplings, polynomially tamed couplings exhibit the following, general expansion:
	\begin{equation}
		\label{TameDC_y_exp_b}
		y(s,a) = \sum_{ \mathbf{m}}\rho_{\mathbf{m}}(e^{-s^i},a^\alpha)   (s^1)^{m_1} \cdots (s^n)^{m_n}\, .
	\end{equation}
	However, they cannot be well-approximated by a monomial as in \eqref{TameDC_mon_tame}; at most, over $\mathcal{E}$, one can upper-bound a polynomially tamed coupling by a monomial as:
	\begin{equation}
		\label{TameDC_pol_tame}
		y(s, a)  \prec  (s^1)^{N_1} \cdots (s^n)^{N_n} \qquad \text{on} \quad \mathcal{E}\ ,
	\end{equation}
	for sufficiently large $N_1, \ldots, N_n$.
	Here the symbol `$\prec$' as to be understood as follows: there exists a parameter $C > 0$ such that $y(s, a)  < C  (s^1)^{N_1} \cdots (s^n)^{N_n}$ holds throughout $\mathcal{E}$.

Clearly, monomially tamed couplings constitute a subset of the more general polynomially tamed couplings.
Furthermore, such definitions can be particularized to a subset $\mathcal{U} \subset \mathcal{E}$, and whether a coupling is monomially or polynomially tamed depends on the specific choice of set $\mathcal{U}$: a coupling can be polynomially tamed $\mathcal{U}$, but monomially tamed on a smaller set within $\mathcal{U}$ or on a different subset within $\mathcal{E}$.

As illustrated in \cite{Grimm:2022sbl}, polynomially tamed couplings are enough to describe all the relevant couplings involving the vector multiplet sector of four-dimensional $\mathcal{N}=2$ Type IIB effective theories, or the respective chiral multiplet sector of the orientifolded $\mathcal{N}=1$ Type IIB effective theories.
The deep reason resides in the fact that the couplings of these sectors descend from the periods of the internal Calabi-Yau manifold.
However, polynomially tamed couplings are expected to be enough to characterize larger families of effective theories.

\subsection{The Distance Conjecture in tame effective theories}
\label{sec:Tame_DC_Tame_and_DC}

In \cite{Grimm:2022sbl}, the tameness of the couplings of the effective theory was employed to show how the predictions of the Distance Conjecture can be realized path independently on some regions of the moduli space, and that the Distance Conjecture may require only a \emph{finite} number of different towers of states.
Since the reasoning that leads to these findings in \cite{Grimm:2022sbl} is similar to the one we will employ in the next section, we will now briefly overview how these features are implied by the tameness of the effective field theory.

To begin with, recall that the main couplings that enter the Distance Conjecture are the factor $e^{-\lambda d(s,a)}$ and the masses of the states within the infinite tower that realize the conjecture.
However, since on different regions of the moduli space different towers may be relevant, for the sake of generality, we consider \emph{several} infinite towers of states.
We denote the masses of the states within these towers as $M_n^{(t)}(s,a)$, with the index $t$ labeling the tower.
Notice that, in principle, the index $t$ might be unbounded, for an effective theory can exhibit infinite candidate towers.
In other words, the set $T$, to which the index $t$ belongs, may be of infinite cardinality.
Moreover, the towers may well be constituted by states with different density: namely, between the some energy scales $\Lambda$ and $\Lambda + \delta \Lambda$, a tower may be constituted by a number $n_1$ of states, and another by a different number $n_2$ of states.

Given the tameness of the underlying effective theory, we may generically infer that $e^{-\lambda d(s,a)}$ and the masses $M_n^{(t)}(s,a)$ need to be tame functions of the axions and saxions.
However, in \cite{Grimm:2022sbl} the following, stronger assumptions for these couplings were adopted:
\begin{itemize}
	\item the masses of all the states constituting the candidate towers $M_n^{(t)}(s,a)$ are polynomially tamed;
	\item $e^{-\lambda d(s,a)}$ is polynomially tamed.
\end{itemize}
These assumptions were motivated in \cite{Grimm:2022sbl} by concrete, consistent effective field theory examples.
It should be noted, however, that the polynomially tame behavior of the masses of candidate towers can be \emph{proven} in some instances. 
For example, this is the case when the masses of the towers can be computed from the periods of the internal manifold, such as the masses of BPS D3-particles or BPS NS5/D5-membrane states of Type IIB four-dimensional effective theories \cite{Grimm:2022sbl}.

Then, in \cite{Grimm:2022sbl} it was proposed to \emph{partition} the patch $\mathcal{E}$ of the moduli space employing the following method.
First, we introduce a partition of the patch $\mathcal{E}$ constituted by some sets $\mathcal{U}_{\mathsf{A}}$. 
Recall that the sets $\mathcal{U}_{\mathsf{A}}$ deliver a partition of $\mathcal{E}$ provided that
\begin{equation}
	\label{TameDC_def-partU}
	\bigcup\limits_{\mathsf{A}}\, \mathcal{U}_{\mathsf{A}} = \mathcal{E}\,, \quad\textrm{and}\quad \mathcal{U}_{\mathsf{A}} \cap \mathcal{U}_{\mathsf{B}} = \varnothing\, , \textrm{ if } {\mathsf{A}}\neq {\mathsf{B}}\,.
\end{equation}
Each subset $\mathcal{U}_{\mathsf{A}}$ of the partition is defined in such a way that, on every subset $\mathcal{U}_{\mathsf{A}}$, $e^{-\lambda d(s,a)}$ displays a \emph{definite} growth.
For instance, on a single subset $\mathcal{U}_{\mathsf{A}}$, $e^{-\lambda d(s,a)}$ may behave as a monomial -- namely, it is monomially tamed, in the language of the previous section; or it may be strictly polynomially tamed (for instance, it may behave as a decreasing exponential in the saxions $s^i$).\footnote{Notice that the possibility to construct such a partition -- as well as the ones that follow -- is due to the tameness of the couplings that induce such a partition.
In fact, since the couplings are tame, one can employ a \emph{monotonicity theorem} \cite{dries_1998} to obtain the partitions defined in this section.}

The partition constituted by the subsets $\mathcal{U}_{\mathsf{A}}$ covers the full patch $\mathcal{E}$, and can be always introduced, provided the tameness of the factor $e^{-\lambda d(s,a)}$.
However, the Distance Conjecture predicts the behavior \eqref{TameDC_Mn} towards boundaries that are located at infinite field distance.
Therefore, among the subsets $\mathcal{U}_{\mathsf{A}}$, we may restrict our attention to the sets that, in \cite{Grimm:2022sbl}, were called `\emph{near-boundary subsets}': these are the subsets $\mathcal{U}_{\mathsf{A}}$ whose closures include the boundary, with the latter assumed to be located at infinite field distance from any other point of the moduli space.

\begin{figure}[thb]
	\centering
	\includegraphics[width=12cm]{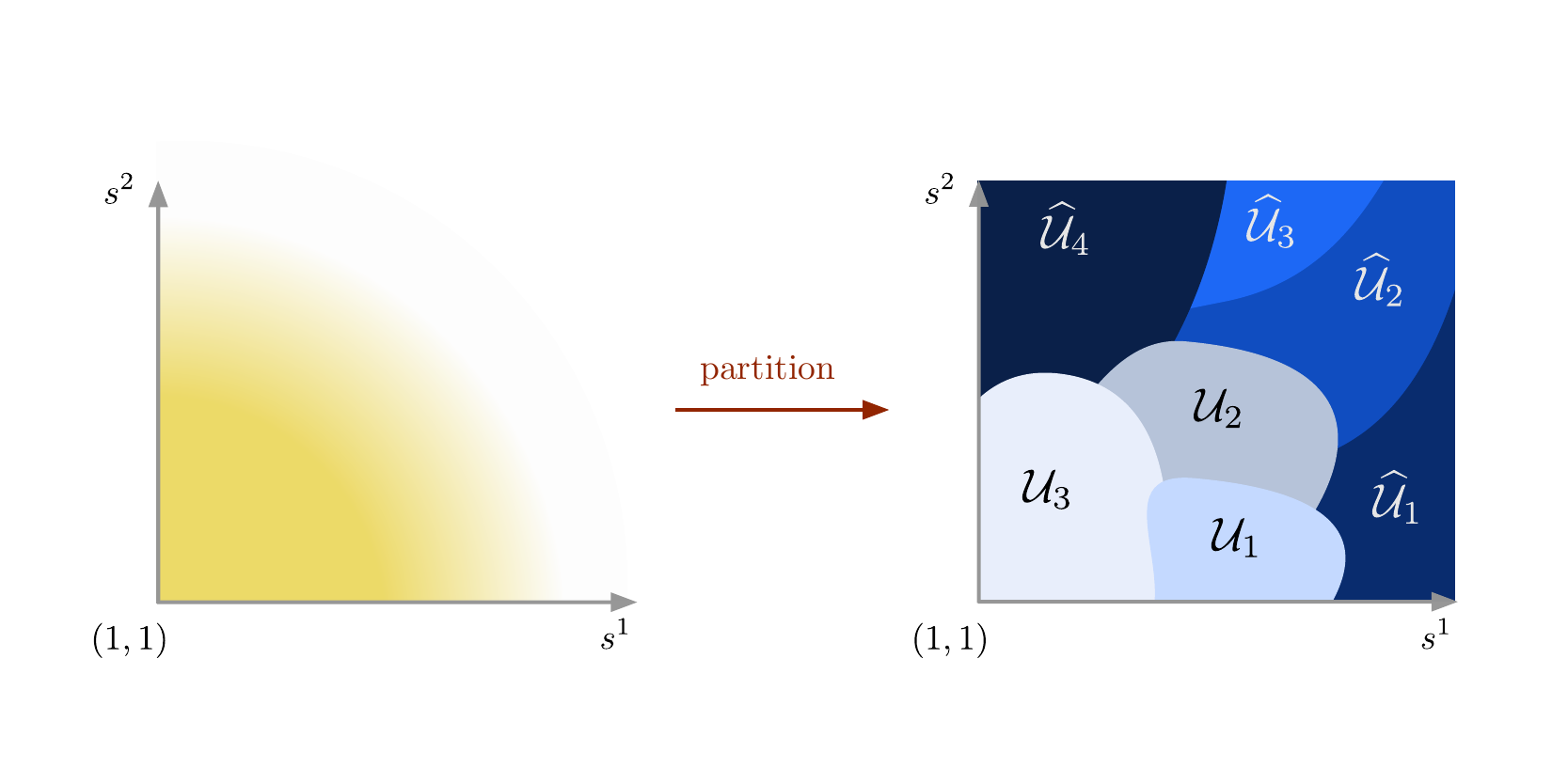}
	\caption{Example of a near-boundary partition of a patch $\mathcal{E}$ of two-dimensional moduli space parametrized by the saxions $s^1 > 1$ and $s^2 > 1$.
		The `near-boundary' subsets $\widehat{\mathcal{U}}_{\mathsf{A}}$, here highlighted with darker colors, are those that extend towards regions where $s^1 \to \infty$ and $s^2 \to \infty$.
		The partition of the patch  $\mathcal{E}$ may however be constitute by other sets, here in lighter colors, that do not include the boundary point.
		\label{Fig:Partition_Mn_a}}
\end{figure}

For the sake of clarity, we will denote with $\widehat{\mathcal{U}}_{\mathsf{A}}$ these near-boundary subsets.\footnote{Concretely, given the boundary point $\varphi_{\rm b}$, these sets are defined as the $\mathcal{U}_{\mathsf{A}}$ such that $\varphi_{\rm b} \in {\rm cl}\,{\mathcal{U}}_{\mathsf{A}}$.}
A pictorial representation of how such a near-boundary subsets may look like can be found in Figure~\ref{Fig:Partition_Mn_a}.
On each of these near-boundary subsets,  since $e^{-\lambda d(s,a)}$ is polynomially tamed, it holds that
\begin{equation}
	\label{TameDC_dist_bound}
	e^{-\lambda d(s,a)}\ \prec  (s^1)^{N^{\mathsf{A}}_1} \cdots (s^n)^{N^{\mathsf{A}}_n} \qquad \text{on any near-boundary}\quad \widehat{\mathcal{U}}_{\mathsf{A}}\,,
\end{equation}
with $N^{\mathsf{A}}_1, \ldots, N^{\mathsf{A}}_n \in \mathbb{Z}_{\leq 0}$, with at least one $N^{\mathsf{A}}_i \in \mathbb{Z}_{< 0}$; clearly, the specific values that bound $e^{-\lambda d(s, a)}$ as in \eqref{TameDC_dist_bound} depend on the near-boundary subset $\widehat{\mathcal{U}}_{\mathsf{A}}$.

Now, if the Distance Conjecture holds true towards any infinite field distance boundary point $\varphi_\text{b}$, it has to hold true, in particular, on each specific near-boundary $\widehat{\mathcal{U}}_{\mathsf{A}}$ separately.
Therefore, among the candidate infinite towers of states, with masses $M_n^{(t)}(s,a)$, there must exist (at least) one tower, the \emph{relevant} tower that realizes the Distance Conjecture labeled by the index $t^r \in T$, such that 
\begin{equation}
	\label{TameDC_Mn_lead}
	M_n^{(t^r)} (s,a) \sim e^{-\lambda d(s,a)} \qquad \text{on \quad $\widehat{\mathcal{U}}_{\mathsf{A}}$\;.}
\end{equation}
It is worth stressing that this relation holds on the full subset $\widehat{\mathcal{U}}_{\mathsf{A}}$, and thus it is true for \emph{any} path within $\widehat{\mathcal{U}}_{\mathsf{A}}$.

Although this will not be discussed further in this work, we recall that in \cite{Grimm:2022sbl} it was further proved that, in order for \eqref{TameDC_Mn_lead} to hold, it is just enough that it holds on specific family of paths that were therein identified as cosmic string solutions.
Therefore, reversing the reasoning, by employing cosmic string solutions one can \emph{test} whether the relation \eqref{TameDC_Mn_lead} holds, and on which near-boundary subset $\widehat{\mathcal{U}}_{\mathsf{A}}$.

In sum, from \eqref{TameDC_Mn_lead}, one can rewrite the near-boundary region spanned by the various subsets $\widehat{\mathcal{U}}_{\mathsf{A}}$ as follows: 
\begin{equation}
	\label{TameDC_Partition_Lead_Tower}
	\bigcup_{\mathsf{A}} \widehat{\mathcal{U}}_{\mathsf{A}} = \left(\bigcup_{k_1} \widehat{\mathcal{U}}^{(1,k_1)} \right) \cup  \left(\bigcup_{k_2} \widehat{\mathcal{U}}^{(2,k_2)} \right) \cup \ldots \cup  \left(\bigcup_{k_M} \widehat{\mathcal{U}}^{(M,k_M)} \right)\,.
\end{equation}
Here the set $\widehat{\mathcal{U}}^{(t^r,k_{t^r})}$ denotes the subset where the tower labeled by the index $t$ is the relevant one that realizes the Distance Conjecture, with $k_{t^r}$ labeling the subsets (among the subsets $\widehat{\mathcal{U}}_{\mathsf{A}}$ that we started with) for which the tower $t^r$ is the relevant one.

Importantly, since the tameness of the effective theory couplings implies that the sum on the left-hand side of \eqref{TameDC_Partition_Lead_Tower} is composed by a \emph{finite} number of subsets $\widehat{\mathcal{U}}_{\mathsf{A}}$, then also the sum on the right-hand side -- that is just a rearrangement of the former -- is composed by a finite number of sets.
As such, only a \emph{finite} number of different towers of states is needed to realize the Distance Conjecture.

\section{Identifying effective-theory breakdown mechanisms via Machine Learning techniques}
\label{sec:ML_and_DC}

In this section we present the founding idea of this work.
Namely, we concretely deliver a method to compute a moduli space partition similar to the one introduced in Section~\ref{sec:Tame_DC_Tame_and_DC}, based on the cutoff of the effective theory, by employing machine learning techniques.
For the sake of conciseness and clarity, we will divide the discussion in two parts: 
first, we will illustrate the properties that the moduli space is expected to exhibit from a top-down viewpoint, building on the discussion of Section~\ref{sec:Tame_DC_Tame_and_DC}; 
then, we show how we can infer these properties from a bottom-up perspective, exploiting the $k$-nearest neighbor and support vector machines algorithms.

\subsection{Top-down viewpoint: an UV-cutoff slicing of the moduli space}
\label{sec:ML_and_DC_UV-slicing}

Across the moduli space, the couplings that characterize the effective field theory change.
Consequently, also the effective field theory cutoff $\Lambda_{\text{\tiny EFT}}$ is expected to be a moduli-dependent function that may drastically change throughout the moduli space.
In this section, the first, generic question that we address is: \emph{which general functional properties of the effective theory cutoff} $\Lambda_{\text{\tiny EFT}}$ \emph{can we infer from a top-down viewpoint?}

First, we assume that the effective theory cutoff $\Lambda_{\text{\tiny EFT}}$ can be related to couplings that appear in the effective field theory, and some examples supporting this assumption will be provided shortly.
As such, the cutoff $\Lambda_{\text{\tiny EFT}}$ is expected to inherit all the general features that any other coupling of the effective theory shares. 
In particular, if the effective field theory is tame, in compliance with the Tameness Conjecture reviewed in Section~\ref{sec:Tame_DC_Tameness}, then $\Lambda_{\text{\tiny EFT}}$ is a tame function of the moduli.
However, as outlined in Section~\ref{sec:Tame_DC_Tame_and_DC}, concrete effective theory exhibit couplings that are not generically tame, but they rather belong to a smaller family of tame functions, namely the one of polynomially tamed functions.
Therefore, we will assume that the effective theory cutoff $\Lambda_{\text{\tiny EFT}}$ is a polynomially tamed function of the moduli as well.

Then, we focus on a near-boundary patch $\mathcal{E}$ of the moduli as defined in \eqref{TameDC_cE}.
The patch $\mathcal{E}$ can then be partitioned following a reasoning similar to the one performed in Section~\ref{sec:Tame_DC_Tame_and_DC}, based on the behavior of the cutoff $\Lambda_{\text{\tiny EFT}}$ understood as function of the axions $a^\alpha$ and the saxions $s^i$.
Specifically, being the cutoff polynomially tamed and bounded as in \eqref{TameDC_pol_tame}, we can partition $\mathcal{E}$ into a \emph{finite} number of smaller subsets $\mathcal{U}_{\mathsf{A}}$ such that, on each $\mathcal{U}_{\mathsf{A}}$, the cutoff $\Lambda_{\text{\tiny EFT}}$ is either monomially tamed, behaving as \eqref{TameDC_mon_tame}, or strictly polynomially tamed, obeying \eqref{TameDC_pol_tame} but not \eqref{TameDC_mon_tame}.
Notice that, with respect to what we did in Section~\ref{sec:Tame_DC_Tame_and_DC}, in this section we will be also interested in the subsets $\mathcal{U}_{\mathsf{A}}$ whose closures do not include the boundary point, for we wish to characterize the behavior of the cutoff throughout the full patch $\mathcal{E}$.

Therefore, analogously to what we did in \eqref{TameDC_Partition_Lead_Tower}, we may rearrange the subsets $\mathcal{U}_{\mathsf{A}}$ of the partition as follows:
on each subset $\mathcal{U}_{\mathsf{A}}$, the cutoff $\Lambda_{\text{\tiny EFT}}$ exhibits a definite growth behavior, and is expected to be related to the behavior of a single effective field theory coupling - say $y^{(r)}$, with $r$ the index labeling the specific coupling.
As an example, $y^{(r)}$ can be identified with the lightest mass $M_{\text{lightest}}^{(r)}$ among the states that could break down the effective description. Alternatively, one can identify $y^{(r)}$ with the species scale \cite{Dvali:2007hz,Dvali:2007wp}, at which quantum gravity effects become non-negligible.
Thus, as in Section~\ref{sec:Tame_DC_Tame_and_DC}, we denote as ${\mathcal{U}}^{(r,k_{r})}$ the subset where the coupling labeled by the index $r$ is the one that gives the cutoff, namely
\begin{equation}
	\label{MLDC_Partition_sets_yr}
	\Lambda_{\text{\tiny EFT}} (s,a) \equiv y^{(r)} (s,a) \qquad \text{on \quad ${\mathcal{U}}^{(y^r,k_{r})}$\;,} 
\end{equation}
and $k_{r}$ labels the different subsets - among the $\mathcal{U}_{\mathsf{A}}$ - where the identification \eqref{MLDC_Partition_sets_yr} holds. 
We can thus rearrange the subsets $\mathcal{U}_{\mathsf{A}}$ partitioning the patch $\mathcal{E}$ as
\begin{equation}
	\label{MLDC_Partition_sets}
	\bigcup_{\mathsf{A}}  {\mathcal{U}}_{\mathsf{A}} = \left(\bigcup_{k_1} {\mathcal{U}}^{(1,k_1)} \right) \cup  \left(\bigcup_{k_2} {\mathcal{U}}^{(2,k_2)} \right) \cup \ldots \cup  \left(\bigcup_{k_M} {\mathcal{U}}^{(M,k_M)} \right)\,.
\end{equation}

\begin{figure}[t]
	\centering
	\includegraphics[width=16cm]{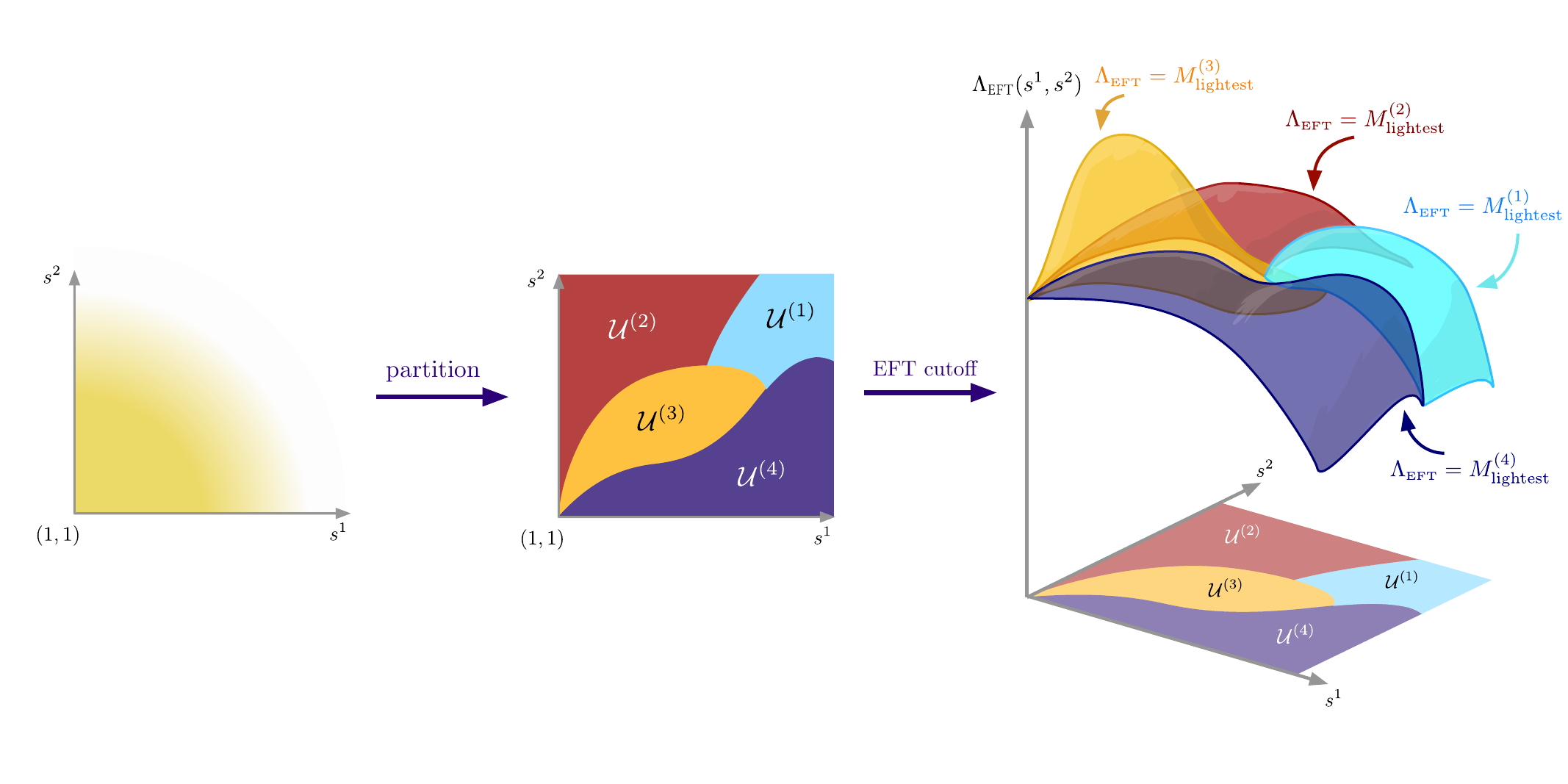}
	\caption{Here is an example of a partition (figure in the center) of a two-dimensional moduli space parametrized by the saxions $s^1 > 1$ and $s^2 > 1$ (depicted on the left). The four subsets of the patch $\mathcal{E}$, here labeled by different colors, denote the unions of subsets $\bigcup_{k_{r}} {\mathcal{U}}^{(y^r,k_{r})} $ where the effective cutoff $\Lambda_{\text{\tiny EFT}}$ is identified with the $y^r$-th coupling. 
	On the right is a schematic depiction of how the effective field theory cutoff $\Lambda_{\text{\tiny EFT}}$ may appear over the partition of the two-dimensional saxionic space.
	On each subset, denoted with different color, the cutoff is related to a different coupling $\Lambda_{\text{\tiny EFT}}$, and it may thus exhibit a different functional behavior.
		\label{Fig:Partition_Cutoff}}
\end{figure}

In order to give an idea of how this partition may look like, a pictorial representation of such a partition of a two-dimensional moduli space, spanned by the saxions $s^1$ and $s^2$, is depicted in Figure~\ref{Fig:Partition_Cutoff}.
The cutoff $\Lambda_{\text{\tiny EFT}}$, which is here a function of the saxions $s^1$ and $s^2$, exhibits a definite growth, or fall-off on each of the subsets plotted in Figure~\ref{Fig:Partition_Cutoff}.
The functional form of the cutoff $\Lambda_{\text{\tiny EFT}}(s^1,s^2)$ is then the union of the various cutoff functions $\Lambda_{\text{\tiny EFT}}(s^1,s^2)$ over the subsets.

The partition induced by the cutoff $\Lambda_{\text{\tiny EFT}}$ may be related to the one introduced in Section~\ref{sec:Tame_DC_Tame_and_DC} in order to realize the Distance Conjecture, at least in the near-boundary subsets $\widehat{\mathcal{U}}_{\mathsf{A}}$.
In fact, the Distance Conjecture may give a recipe in order to identify the cutoff of an effective theory that admits a quantum gravity UV completion.
Indeed, within each of the near-boundary subsets $\widehat{\mathcal{U}}_{\mathsf{A}}$ an infinite tower of states becoming light, and the maximal cutoff that allows the effective field theory to be valid in order to avoid the inclusion of the entirety of the tower is
\begin{equation}
	\label{MLDC_Lambda}
	\Lambda_{\text{\tiny EFT}} = M_{\text{lightest}}^{(t^r)} \qquad \text{on the near-boundary sets \quad $\widehat{\mathcal{U}}_{\mathsf{A}}$,}
\end{equation}
where $M_{\text{lightest}}^{(t^r)}$ denotes the lightest among the masses $M_n^{(t^r)} (s,a)$ of the states of the infinite tower (labeled by the index $t^r$) that realizes the Distance Conjecture as in \eqref{TameDC_Mn_lead} over $\widehat{\mathcal{U}}_{\mathsf{A}}$.

Therefore, the near-boundary sets $\widehat{\mathcal{U}}_{\mathsf{A}}$ of the partition $\{ \mathcal{U}_{\mathsf{A}} \}$ induced by the cutoff are expected to be similar -- if not equal -- to those that realize the partition as in \eqref{TameDC_Mn_lead}.
For example, in these subsets the cutoff $\Lambda_{\text{\tiny EFT}}$ may be identified with the mass of the lightest among the Kaluza-Klein modes or winding modes; or, away from decompactification limits, $\Lambda_{\text{\tiny EFT}}$ may be identified with an emergent string scale \cite{Lee:2018urn,Lee:2018spm,Lee:2019tst,Lee:2019xtm,Lee:2019wij,Klaewer:2020lfg}.

Alternatively, other brane states may deliver the effective theory cutoff within near-boundary subsets.
For instance, in four-dimensional Type IIB effective theories, the Distance Conjecture may be realized by infinite towers of D3-particles \cite{Grimm:2018ohb,Grimm:2018cpv,Gendler:2020dfp,Palti:2021ubp} or D5/NS5-membrane states \cite{Lanza:2020qmt,Grimm:2022sbl}; then, the cutoff may be given by the lowest among the masses of these states, and these were proven to be polynomially tamed in \cite{Grimm:2022sbl}, thus motivating the hypothesis employed at the beginning of this section.

For the sets of the partition $\{ \mathcal{U}_{\mathsf{A}} \}$ that do not include the boundary point the identification of the cutoff might be more subtle. Therein, the cutoff might be still given by the lightest mass of an infinite tower of states (that, however, does not become massless in said subset).
More generally, some other state that is not necessarily part of an infinite tower can be the first one to be encountered when probing higher energies, and thus its mass may be employed as effective field theory cutoff $\Lambda_{\text{\tiny EFT}}$ so as to avoid its inclusion in the effective theory that we started with.

\subsection{Bottom-up viewpoint I: partitioning the moduli space via a \texorpdfstring{$k$}{k}-nearest neighbor algorithm}
\label{sec:ML_and_DC_k-nearest}

We will now illustrate how the partition of the moduli space that follows the recipe in \eqref{MLDC_Lambda} can be uncovered from a bottom-up perspective.
Namely, consider an effective field theory with a nontrivial moduli space, defined over a patch $\mathcal{E}$ that is parametrized by the coordinates $\varphi^A$.
Here, we shall address the following question: \emph{given an arbitrary point $\varphi^A_0 \in \mathcal{E}$, what is the type of states that would `likely' break down the effective field theory there?}

Knowing the answer to this question for a sufficient number of points would allow one to \emph{reconstruct} the partition dictated by \eqref{MLDC_Lambda}, and eventually infer some key properties about the UV-completion of the effective theory under investigation. 
In this section, we will first answer the question by exploiting the $k$-nearest neighbor algorithm. This is a supervised machine learning algorithm that is commonly used in Data Science.
However, before showing how to use such an algorithm, we first need to specify the data that one should minimally feed to the algorithm to make it work.

\noindent\textbf{The data.} Consider a generic effective theory with a nontrivial moduli space, and assume that we are able to measure some physical quantities that the effective theory describes.
Specifically, let us assume that, with appropriate measurements carried at some spacetime point $x^\mu_1$, we are able to infer the vev's of the moduli fields at that point, $\varphi^A(x_1) \equiv \varphi^A_1$.
In other words, the vev's $\varphi^A_1$ are those that reproduce the value of the various physical quantities that we measure.

Assume that we know that our effective theory is valid up to some energy scale $E \leq \Lambda_1$.
However, we wish to get some \emph{minimal} information about the UV-completion of the effective theory.
Therefore, let us assume that we can increase the energy scales $E$ that are probed beyond $\Lambda_1$.
At some energy scale $\Lambda_1' \geq \Lambda_1$ a new state will be detected, and the effective field theory we started with may not be enough to describe its dynamics: we say that this state is the source of the breakdown of the effective theory, and its mass $m_1 \simeq \Lambda_1'$ may provide the cutoff of the effective theory that we started with, and we set $\Lambda_{\text{\tiny EFT}}(\varphi_1) = m_1$.
Let us denote the type of this state as `\emph{Type I}'. 
For instance, the type may be the collection of the quantum numbers that specify the state, or it can specify the microscopic origin of the state, for instance the type of brane which it originates from.

Now, let us assume that we can move our probing instruments to some other location, at some different coordinates $x^\mu_2$.
The vev's of the moduli at the spacetime point $x^\mu_2$ might be different from the one at the point $x^\mu_1$.\footnote{For instance, the presence of dynamical objects such as black holes \cite{toappear_withJeroen}, cosmic strings \cite{cstring,Lanza:2021udy} and domain walls \cite{Lanza:2020qmt} may be the source of a nontrivial change of the vev's of the moduli throughout spacetime.}
Following the same procedure as the one carried at the spacetime point $x^\mu_1$, let us increase the energy scales probed until we find a new state that would thus break down the effective theory that we started with. 
As above, the mass of this state would tell us the cutoff of the effective field theory at the moduli space point $\varphi^A_2$, as $\Lambda_{\text{\tiny EFT}}(\varphi_2) = m_2$.
This state may be the same that we previously found at $x^\mu_1$, namely a `\emph{Type I}' state; or it could be a different kind of state -- say, of `\emph{Type II}'.

We assume that this procedure can be replicated at several, different spacetime points, corresponding to different points in the moduli space.
The set of the moduli space points $\{\varphi^A_k\}$ probed in this way, along with the types $\text{Type}\, J_{k}$ of the states that first break down the effective theory at the point $\varphi^A_k$ constitute the data that we will be focusing on and which we will feed to the machine learning algorithm:
\begin{equation}
	\label{MLDC_data}
	\text{Data} = \{ (\varphi^A_{k},\, \text{Type}\, J_{k}) \}\,.
\end{equation}

Clearly, at the moment, we do not have access to such a phenomenological knowledge.
In order to compensate to this lack, and in order to show how the algorithm of the following section works, we will rather generate the data \eqref{MLDC_data} by exploiting the known microscopic description of the effective theory.
In Data Science language, these sorts of data that are not obtained from external sources, but they are rather artificially crafted, are oftentimes called `\emph{synthetic data}'.

\begin{figure}[thb]
	\centering
	\includegraphics[width=11cm]{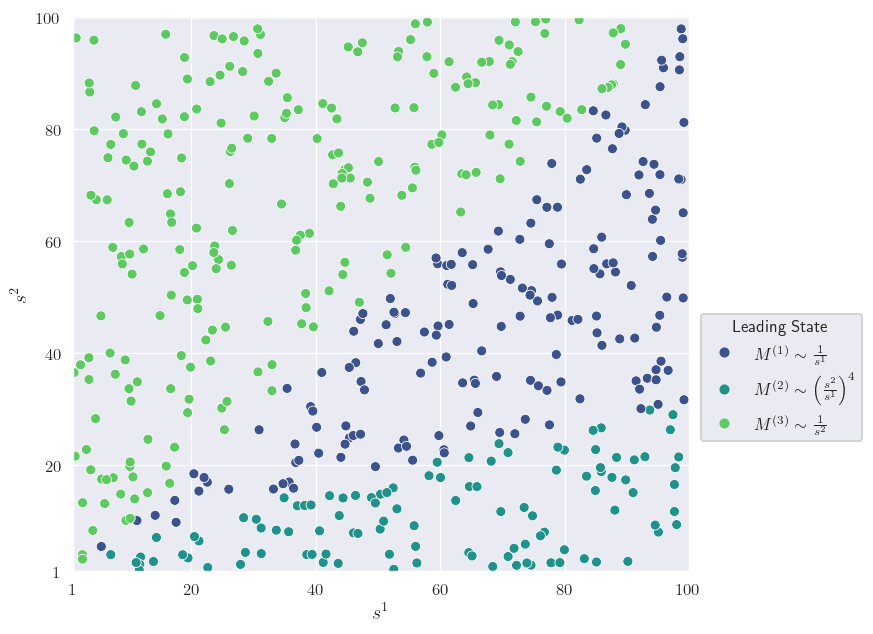}
	\caption{An example of data for a two-dimensional field space, spanned by the saxions $s^1$ and $s^2$. 
		Each point represents one measurement, and the color labels the kind of state that breaks down the effective theory.
		The states are here assumed to scale, in terms of the saxions, as $\sim \frac{1}{s^2}$ (green dots), $\sim \frac{1}{s^1}$ (blue dots), $\sim \left(\frac{s^2}{s^1}\right)^4$ (teal dots).
		\label{Fig:Example_Data}}
\end{figure}

\begin{figure}[thb]
	\centering
	\includegraphics[width=11cm]{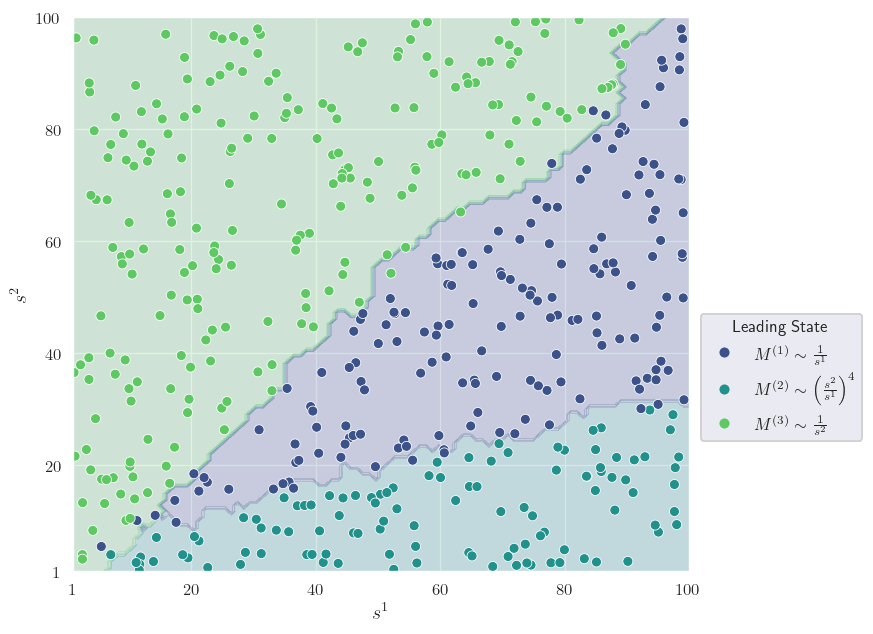}
	\caption{The application of the $k$-nearest neighbor algorithm to the data plotted in Figure~\ref{Fig:Example_Data} leads the regions here depicted, with the color scheme following the same as for the towers in Figure~\ref{Fig:Example_Data}.
		The algorithm has an accuracy of $0.97$ on the training test set, and an accuracy of $0.99$ on the test set.
		\label{Fig:Example_Partition}}
\end{figure}

In Figure~\ref{Fig:Example_Data} is a toy-example of such a synthetic data set, which has been obtained as follows.\footnote{The plots in Figure~\ref{Fig:Example_Data} and Figure~\ref{Fig:Example_Partition}, and the following are obtained from a \texttt{python} code, employing the \texttt{pandas} and \texttt{matplotlib} libraries. The former is particularly convenient for dealing and generating large data sets, while the second allows for the graphical visualization of data.} 
Consider an effective theory, which we assumed to be endowed with a nontrivial moduli space.
Within its moduli space, we focus on a two-dimensional subset thereof, that is spanned by two saxions, $s^1$ and $s^2$, while fixing the other moduli to some given vev's. 
We assume that the states that could break down the effective theory could belong to three species, whose masses are as $M^{(1)} = \frac{1}{s^1}$, $M^{(2)} = \left(\frac{s^2}{s^1}\right)^4$, $M^{(3)} = \frac{1}{s^2}$.
For instance, they might be the lightest states of some towers of infinite states, scaling in this fashion in the saxions $s^1$ and $s^2$.

Then, we randomly generate 500 moduli space points $(s^1, s^2)$, in order to \emph{simulate} some phenomenological measurements probing those moduli space points.
For each point so generated, we determine which is the relevant state that leads to the breakdown of the effective field theory: namely, we determine which is the lightest among the states with masses $M^{(1)}$, $M^{(2)}$ and $M^{(3)}$.
The different colors of the points in Figure~\ref{Fig:Example_Data} tell which of the three species deliver the leading state breaking down the effective theory. 

\noindent\textbf{The $k$-nearest neighbor algorithm.} Datasets like the one depicted in Figure~\ref{Fig:Example_Data} constitute the sources upon which we can apply machine learning methods.
As anticipated earlier, we will first employ the $k$-nearest neighbor algorithm.
Here we will highlight some of the basic features of the algorithm, and we refer to Appendix~\ref{sec:k-near_review} for a more detailed review thereof.

Preliminarily, it is worth stressing that the question at hand can be regarded as a \emph{classification} problem that requires \emph{supervised} machine learning techniques.
Indeed, in terms of the new terminology introduced, the question, mentioned at the beginning of this section, that we wish to address can be reformulated as follows: given a point $\varphi^A_{\text{\tiny new}}$ of the moduli space that is \emph{not} contained in the dataset \eqref{MLDC_data} that we started with, what is the \emph{Type} $J_{\text{\tiny new}}$ of state that will likely first break down the effective theory at $\varphi^A_{\text{\tiny new}}$?
Therefore, this is a \emph{classification} problem because we wish to tell to which category (i.e.~the Type $J_{\text{\tiny new}}$ of the breakdown state) the new variable (i.e.~the point $\varphi^A_{\text{\tiny new}}$) is associated with.
As explained in Section~\ref{sec:ML_and_DC_UV-slicing}, the tameness of the effective theory guarantees that the Type $J_{\text{\tiny new}}$ belong to a \emph{finite} set.
Furthermore, the problem requires a \emph{supervised} approach since we know the kind of `\emph{target}' that we wish to predict -- i.e.~the Type $J_{\text{\tiny new}}$ of the breakdown state.

The $k$-nearest neighbor algorithm is a supervised machine learning algorithm that addresses precisely such classification problems via the following procedure.
First, as any other machine algorithm, it requires a training dataset, and a test dataset.
In the following, we will devote $80\%$ of the data in \eqref{MLDC_data} to the training set, and the remaining $20\%$ to the test set.

Now, consider a moduli space point $\varphi^A_{\text{\tiny new}}$, not included in the training part of the dataset \eqref{MLDC_data}.
The algorithm computes the Euclidean distance from the point $\varphi^A_{\text{\tiny new}}$ to all the points in the training set, and picks the closest $k$ points.
Then, among such $k$ neighbors, the algorithm determines what is the most recurring Type $J_{k}$ of states that break down the effective description.
The most common type will be the Type $J_{\text{\tiny new}}$ assigned to $\varphi^A_{\text{\tiny new}}$.

We remark that the choice of $k$ is arbitrary, and one could tune $k$ in a such a way that the algorithm delivers the best performance on the test set.
Typically, one should expect that, the higher the value of $k$, the more stable the results of the algorithm are.
However, a too great value of $k$ might just trivialize the algorithm and deliver wrong predictions, and a compromise choice has to be taken.

In the following, we will employ such an algorithm to plot `\emph{decision regions}'.
In order to obtain the latter we consider a \emph{grid}, composed of several hundred points that cover the subset of the moduli space that we will be focusing on.
Then, we apply the $k$-nearest neighbor algorithm for each point of the grid, with the decision based on the training set, and the decision regions are plotted joining the points of the grid belonging to the same type.
The decision regions so obtained portray the approximate shape of the subsets that constitute the UV-cutoff-induced partition.

An example of output delivered by the $k$-nearest neighbor algorithm is contained in Figure~\ref{Fig:Example_Partition}, where we applied the algorithm to the synthetic data shown in Figure~\ref{Fig:Example_Data}.

\subsection{Bottom-up viewpoint II: obtaining decision boundaries with Support Vector Machines}
\label{sec:ML_and_DC_SVM}

The $k$-nearest neighbor algorithm allows for obtaining a partition of the moduli space in general. However, due to the very nature of the algorithm, the decision boundaries that separate the subsets of the partition might be quite irregular -- an issue that becomes more and more severe as the value of nearest neighbors $k$ is lowered.
As a result, it might be hard to determine the shape of the decision boundaries, as in Figure~\ref{Fig:Example_Partition}.
In order to sharpen the prediction of the decision boundaries some other, more sophisticate machine learning algorithms may be employed, such as the `\emph{support vector machine algorithms}', one of which we will now discuss.

For simplicity, we will focus on the case in which the masses of the states $M^{(r)}$ that could break down the effective theory as in \eqref{MLDC_Lambda} are all monomially tamed, behaving as in \eqref{TameDC_mon_tame} in the entire patch $\mathcal{E}$ of the moduli space.
Such masses behave as monomials in the saxions as the boundaries of the moduli space patch $\mathcal{E}$ are approached, and $\log M^{(r)}$ is well-approximated by a \emph{linear} combination of the logarithms of the saxions:
\begin{equation}
	\label{MLDC_Mtr_montam}
	\log M^{(r)} \sim \sum\limits_{i = 1}^n  k_i^{(r)} \log s^i  \qquad \text{on} \quad \mathcal{E}\ .
\end{equation}
For instance, this is the case of the states that break down the effective description in Figure~\ref{Fig:Example_Data}, where the states are actual monomials of the saxions $s^1$ and $s^2$.

With this simplifying hypothesis in force, the characterization of the decision boundary becomes simpler. 
Indeed, consider two neighboring, near-boundary subsets of $\mathcal{E}$, $\widehat{\mathcal{U}}^{(r^1)}$ and $\widehat{\mathcal{U}}^{(r^2)}$: we assume that in the former the state with mass $M^{(r^1)}$ is the one that breaks down the effective theory, serving as the effective theory cutoff, while the state with mass $M^{(r^2)}$ serves as cutoff for the second subset $\widehat{\mathcal{U}}^{(r^2)}$.
Clearly, it needs to hold that $M^{(r^1)} \leq M^{(r^2)}$ in $\widehat{\mathcal{U}}^{(r^1)}$, whereas $M^{(r^2)} \leq M^{(r^1)}$ in $\widehat{\mathcal{U}}^{(r^2)}$.
The decision boundary that separates the subsets $\widehat{\mathcal{U}}^{(r^1)}$ and $\widehat{\mathcal{U}}^{(r^2)}$ is thus the locus on which $M^{(r^1)} = M^{(r^2)}$.
However, since we assumed that the behavior of the masses $M^{(r^1)}$ and $M^{(r^2)}$ obeys \eqref{MLDC_Mtr_montam}, as the boundaries are approached, the locus where $M^{(r^1)} = M^{(r^2)}$ resembles an \emph{hyperplane} in the $\log s^i$ variables.

Clearly, the procedure illustrated can be repeated for any other pair of neighboring near-boundary sets, which will still be approximately separated by boundary decision hyperplanes, due to the monomially tamed behavior \eqref{MLDC_Mtr_montam}.

\begin{figure}[thb]
	\centering
	\includegraphics[width=11cm]{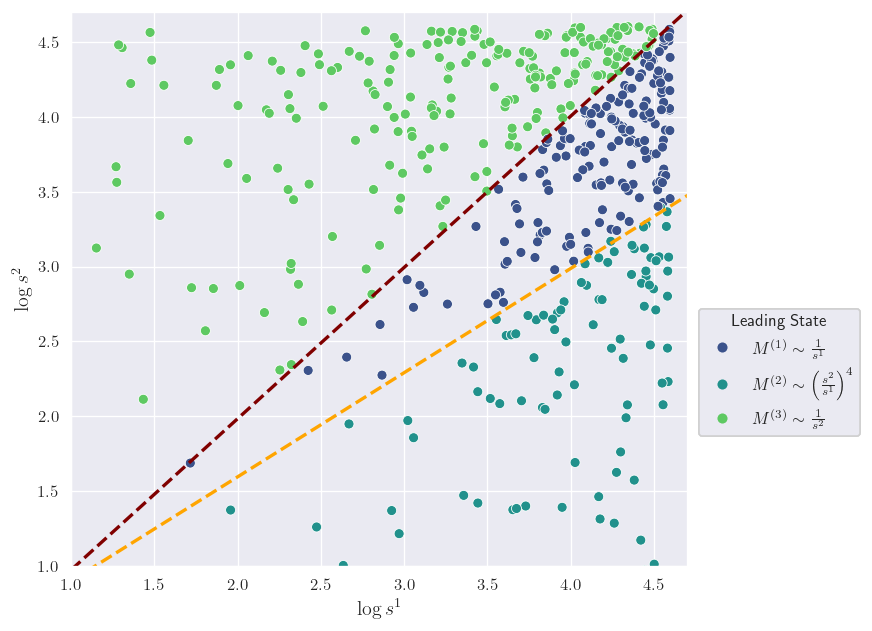}
	\caption{Here is plotted the application of the linear support vector machine algorithm to the dataset in Figure~\ref{Fig:Example_Data}.
	The algorithm produces the two dashed line plotted above, which separate the three subsets of the two-dimensional saxion space where each state is relevant.
	The slope of the upper, darker line is approximately $\simeq 0.99$, while the slope of the lower, lighter line is $\simeq 0.74$, very close to the expected values of $1$ and $0.75$, respectively.
	\label{Fig:Example_SVM}}
\end{figure}

Support vector machines are machine learning algorithms that aim at separating the data according to the category they belong.
The setup and the structure of the data on which the algorithm is based on are the same as the ones that we introduced in the previous section for the $k$-nearest neighbor algorithm: the dataset that we start with is as in \eqref{MLDC_data} and is composed by a set of moduli space points $\varphi^A_k$, alongside with the type $\text{Type} J_k$ of state that breaks down the effective theory at $\varphi^A_k$. 
For example, the dataset depicted in Figure~\ref{Fig:Example_Data} may well serve as the starting point for support vector machine algorithms.

Due to the hypothesis \eqref{MLDC_Mtr_montam}, for our scopes it is enough to employ \emph{linear} support vector machine algorithms.
After feeding data of the form as in \eqref{MLDC_data}, the algorithm searches for (portions of) hyperplanes that best separate the dataset according to their categorical value.
Specifically, in our case, the algorithm can find hyperplanes in the $\log s^i$ variables that best separate the moduli patch of interest according to the type of state that breaks down the effective theory.
An overview of how the linear support vector machine algorithm is contained in Appendix~\ref{sec:LSMVMreview}.
As an example in Figure~\ref{Fig:Example_SVM} is the application of the linear support vector machine algorithm to the dataset depicted of Figure~\ref{Fig:Example_Data}. 
The algorithm delivers the boundaries that best separate the three distinct subsets where each effective theory-breakdown state is relevant. 

Although the linear support vector machine algorithms allow for having full control of the boundary decision, they clearly have some limitations.  
Firstly, it is worth stressing that, even with the hypothesis \eqref{MLDC_Mtr_montam} in force, the decision boundaries are \emph{asymptotically} hyperplanes: for small values of the saxions, far away from the boundaries of $\mathcal{E}$, although \eqref{MLDC_Mtr_montam} holds, $\log M^{(r)}$ might not be approximated by a linear combination of the logarithms of the saxions sufficiently well.
Moreover, if the couplings are not monomially tamed but, rather polyonomially tamed, then the decision boundaries are not hyperplanes in general, even in the near-boundary regime.
Nonlinear version of support vector machine algorithms have also been introduced: they exploit a `\emph{kernel trick}' consisting of artificially recasting the problem as a higher-dimensional one, where the data can be separated by hyperplanes.
The application of nonlinear support vector machines to the UV-cutoff slicing of the moduli space, which are necessary when the couplings breaking down the effective theory are not monomially tamed, is left for future work.

\section{A concrete example: a toroidal orbifold compactification}
\label{sec:Ex_Tor}

This section is devoted to showing how the ideas explained in Section~\ref{sec:ML_and_DC} apply to a concrete effective field theory.
The effective field theory that we consider is a four-dimensional $\mathcal{N}=2$ supergravity effective field theory obtained after compactifying the ten-dimensional Type IIB string theory over a Calabi-Yau manifold, and we will choose the latter to be the toroidal orbifold $T^6/(\mathbb{Z}_2 \times \mathbb{Z}_2')$ considered in \cite{Blumenhagen:2003vr}.
The reasons for focusing on this model rely on the simplicity of the relation of the moduli with respect to the geometric data, and on the clear identification of the candidate infinite towers of states that can break down the effective description.

Let us review briefly the relevant features exhibited by the the four-dimensional compactification that we consider.
Preliminarily, we recall that the ten-dimensional Type IIB effective theory that we start with, in the conventions of \cite{Ibanez:2012zz}, is
\begin{equation}
	\begin{aligned}
		\label{ExTor_SIIB}
		S_{\text{\tiny EFT}}^{(10)} = &\frac{1}{2 \kappa^2_{10}}\int d^{10}x\, \sqrt{-g} \Bigg[ e^{-2\phi} \left( R + 4 \partial \phi \cdot \partial \phi - \frac12 |H_3|^2  \right) - \frac12 |F_1|^2 - \frac12 |\tilde{F}_3|^2 - \frac14 |\tilde{F}_5|^2  \Bigg] 
		\\
		&\qquad\, - \frac{1}{4 \kappa^2_{10}} \int C_4 \wedge H_3 \wedge F_3\,,
	\end{aligned}
\end{equation}
where $2 \kappa^2_{10} =  2\pi / \ell_\text{s}^8$, with the string length $\ell_\text{s}^2 = 2\pi \alpha'$, and string mass $M_\text{s} = \frac{1}{\ell_\text{s}}$.
Here, $\phi$ denotes the ten-dimensional dilaton, related to the string coupling as $g_\text{s} = e^\phi$, $H_3 = {\rm d} B_2$ and $F_1$, $\tilde{F}_3$ and $\tilde{F}_5$ being the gauge invariant Ramond-Ramond field strengths.

We will consider the following ansatz for the ten-dimensional spacetime metric:
\begin{equation}
	\label{ExTor_dsans}
	{\rm d}s^2_{10D} = e^{2A} {\rm d}s^2_{4D} + {\rm d}s^2_{Y}\,,
\end{equation}
with $Y = T^6/(\mathbb{Z}_2 \times \mathbb{Z}_2')$ being the six-dimensional internal manifold.
We shall choose the dimensionless warp factor $e^{2A}$ in such a way that the effective theory \eqref{ExTor_SIIB}, once recast to the ten-dimensional Einstein frame, and then reduced over the metric \eqref{ExTor_dsans}, delivers a four-dimensional effective theory that is expressed in the four-dimensional Einstein frame:
\begin{equation}
	\label{ExTor_e2A}
	e^{2A} = \frac{M_\text{P}^2 \ell_\text{s}^8}{4 \pi V_Y}\,,
\end{equation}
where $V_Y$ is the volume of the internal manifold $Y$ expressed in the Einstein frame.

\noindent\textbf{On the internal $T^6/(\mathbb{Z}_2 \times \mathbb{Z}_2')$ orbifold.} Let us denote with $z^i$, with $i = 1, 2, 3$ the three holomorphic coordinates of the internal torus, and with $\hat{x}^j$, $j = 1,\ldots, 6$ the internal, real coordinates. In the following, it will be convenient to take the coordinates $\hat{x}^j$ dimensionless, and such that they span circles of unitary length.
As such, the holomorphic coordinates $z^i$ and the real coordinates $\hat{x}^j$ so defined can be related as
\begin{equation}
	\label{ExTor_dzi}
	{\rm d} z^i = 2\pi \left( R_{2i - 1} {\rm d} \hat{x}^{2i - 1} + \im R_{2i} {\rm d} \hat{x}^{2i} \right)
\end{equation}
where $R_i$ is the $i$-th internal radius, expressed in Einstein frame. 

The orbifold $\mathbb{Z}_2$ and $\mathbb{Z}_2'$ operations act on the internal coordinates as follows \cite{Blumenhagen:2003vr}
\begin{equation}
	\label{ExTor_Z2orb}
	\theta : \begin{cases} z^1 \to - z^1 \\  z^2 \to - z^2 \\ z^3 \to z^3 \end{cases} \qquad , \qquad \theta' : \begin{cases} z^1 \to z^1 \\  z^2 \to - z^2 \\ z^3 \to -z^3 \end{cases} \,.
\end{equation}
A choice of symplectic basis of three-forms $\{\alpha_I, \beta^J\}$, obeying $\int_Y \alpha_I \wedge \beta^J = \delta_I^J$ and that is invariant under the orbifold action \eqref{ExTor_Z2orb} is
\begin{equation}
	\label{ExTor_sympl_basis}
	\begin{aligned}
		&\alpha_0 = {\rm d} \hat{x}^1 \wedge  {\rm d} \hat{x}^3  \wedge {\rm d} \hat{x}^5 \qquad &, \qquad  & \alpha_1 = {\rm d} \hat{x}^2 \wedge  {\rm d} \hat{x}^3  \wedge {\rm d} \hat{x}^5 \qquad ,
		\\
		&\alpha_2 = {\rm d} \hat{x}^1 \wedge  {\rm d} \hat{x}^4  \wedge {\rm d} \hat{x}^5 \qquad &, \qquad  & \alpha_3 = {\rm d} \hat{x}^1 \wedge  {\rm d} \hat{x}^3  \wedge {\rm d} \hat{x}^6 \qquad ,
		\\
		&\beta^0 = - {\rm d} \hat{x}^2 \wedge  {\rm d} \hat{x}^4  \wedge {\rm d} \hat{x}^6 \qquad &, \qquad  & \beta^1 = {\rm d} \hat{x}^1 \wedge  {\rm d} \hat{x}^4  \wedge {\rm d} \hat{x}^6 \qquad ,
		\\
		&\beta^2 = {\rm d} \hat{x}^1 \wedge  {\rm d} \hat{x}^3  \wedge {\rm d} \hat{x}^6 \qquad &, \qquad  & \beta^3 = {\rm d} \hat{x}^2 \wedge  {\rm d} \hat{x}^4  \wedge {\rm d} \hat{x}^5 \qquad .
	\end{aligned}
\end{equation}

\noindent\textbf{The moduli space.} The moduli space of the four-dimensional theory is composed by three sectors. Here, we shall focus on the saxionic part of these sectors only. Firstly, the ten-dimensional dilaton $\phi$ appears in the four-dimensional action as modulus. 
The saxion, defined out of the ten-dimensional dilaton, $s_0 = e^{-\phi} = \frac{1}{g_\text{s}}$ is assumed to be large, so as the string coupling is small. A second sector is formed by the K\"ahler moduli $v_i$, which are most readily associated with the areas of the internal two-dimensional tori as
\begin{equation}
	\label{ExTor_Kahl_mod}
	v_i = \left(\frac{2\pi}{\ell_\text{s}}\right)^2 R_{2i-1} R_{2i}\,.
\end{equation}
The third sector, composed by the tori complex structure moduli, is slightly more involved to be identified.
The first step is to recast the Calabi-Yau unique holomorphic three-form $\Omega = {\rm d} z^1 \wedge {\rm d} z^2 \wedge {\rm d} z^3$ in terms of the symplectic basis \eqref{ExTor_sympl_basis} as follows:
\begin{equation}
	\label{ExTor_Omega_sympl}
	\Omega = \ell_\text{s}^3 \left(X^I \alpha_I - \mathcal{F}_I(X) \beta^I\right)\,.
\end{equation}
Here, $\mathcal{F}_I(X)$ denote the derivatives of the holomorphic prepotential $\mathcal{F}(X) = - \frac{X^1 X^2 X^3}{X^0}$ with respect to the coordinates $X^I$. Then, inserting \eqref{ExTor_dzi} and \eqref{ExTor_sympl_basis} in the identification \eqref{ExTor_Omega_sympl}, one obtains how the coordinates $X^I$ and $\mathcal{F}_I(X)$ are expressed in terms of the geometric data. Then, the complex structure saxions $s^i$ are identified in terms of the internal $S^1$ radii as
\begin{equation}
	\label{ExTor_cs_mod}
	s^i = {\rm Im} \frac{X^i}{X^0} = \frac{R_{2i}}{R_{2i-1}}\,.
\end{equation}

For future reference, we report here how the radii of the internal $S^1$ are expressed in terms of the K\"ahler and complex structure moduli by inverting \eqref{ExTor_Kahl_mod} and \eqref{ExTor_cs_mod}
\begin{equation}
	\label{ExTor_Ri}
	R_{2i - 1}^2 = \left(\frac{\ell_\text{s}}{2\pi}\right)^2 \frac{v_i}{s^i}\,, \qquad R_{2i}^2 = \left(\frac{\ell_\text{s}}{2\pi}\right)^2 v_i s^i\,, 
\end{equation}
and we note the Einstein-frame internal volume is $V_Y = \ell_\text{s}^6 v_1 v_2 v_3$.

\subsection{Candidate effective theory-breakdown states}
\label{sec:Ex_Tor_scales}

The four-dimensional effective field theories obtained from the compactification of Type IIB string theory over the aforementioned $T^6/(\mathbb{Z}_2 \times \mathbb{Z}_2')$ orbifold may be broken down by different sources, as the energy scales that are probed are increased.
Indeed, the microscopic description predicts the existence of several infinite towers of states that can break down the effective theory, which we now enlist.

\noindent\textbf{Tower Ia: Kaluza-Klein Modes.} The Kaluza-Klein modes are prime candidates that could break down effective field theories stemming from a compactification of a higher dimensional theory. 
Moreover, as outlined in \cite{Ooguri:2006in}, since they come in infinite number, they could well be the infinite states that realize the Distance Conjecture. 
Here we will be specifically interested in the \emph{scalar} Kaluza-Klein modes.  
They can originate, for instance, from the excitations of the ten-dimensional dilaton $\phi$ appearing in \eqref{ExTor_SIIB} along some of the internal dimension. 
For the toroidal model at hand, one can identify six different types of Kaluza-Klein mode, according to the $S^1$ along which the excitation occurs.

In order to compute the masses of the dilation-originated Kaluza-Klein modes one can start with the ten-dimensional action \eqref{ExTor_SIIB}, switch to the Einstein frame, expand the dilaton in Fourier component along the chosen internal $S^1$, and then compactify the theory using the metric ansatz \eqref{ExTor_dsans}, with the warp factor chosen as in \eqref{ExTor_e2A}.
Finally, one gets the following, four-dimensional masses for the six different types of Kaluza-Klein modes:
\begin{equation}
	\label{ExTor_mKK}
	m^{\text{KK}, i}_{k} = \frac{2 \pi k}{R_i} e^A\,,
\end{equation}
where $k \in \mathbb{N}$ denotes the excitation level, and $R_i$ is the Einstein-frame radius of the $S^1$ delivering the excitation.
Exploiting the relations \eqref{ExTor_e2A} and \eqref{ExTor_Ri} we can re-express the Kaluza-Klein masses in terms of the moduli as
\begin{equation}
	\label{ExTor_mKK_b}
	m^{\text{KK}, 2i-1}_{k} = M_\text{P} (2\pi k) \sqrt{\frac{\pi}{v_1 v_2 v_3}} \frac{s^i}{v_i}\,, \qquad m^{\text{KK}, 2i}_{k} = M_\text{P} (2\pi k) \sqrt{\frac{\pi}{v_1 v_2 v_3}} \frac{1}{ s^i v_i}\,.
\end{equation}

\noindent\textbf{Tower Ib: Winding Modes.} A second candidate of tower of states that could break down any effective theory that originates from a compactification is the tower of winding modes. 
This tower is dual of the Kaluza-Klein modes tower discussed above.
The masses for the winding modes can be most readily obtained from the ones of the Kaluza-Klein modes in \eqref{ExTor_mKK} upon exchanging $R_i \to \alpha'/R_i$:
\begin{equation}
	\label{ExTor_mw}
	m^{\text{w}, i}_{k} = \frac{2 \pi k}{\alpha'} e^A R_i \,,
\end{equation}
which can be recast, in terms of the moduli, as
\begin{equation}
	\label{ExTor_mw_b}
	m^{\text{w}, 2i-1}_{k} = M_\text{P} k  \sqrt{\frac{\pi}{v_1 v_2 v_3}} \frac{v_i}{s^i}\,, \qquad m^{\text{w}, 2i}_{k} = M_\text{P} k \sqrt{\frac{\pi}{v_1 v_2 v_3}}  s^i v_i\,.
\end{equation}

Kaluza-Klein and winding modes may be the source of the effective description breakdown for any theory that stems from dimensional reductions.
However, let us now specialize to potential breakdown sources that are peculiar to Type IIB effective theories.
These may originate from the brane content that characterizes the ten-dimensional Type IIB string theory.

\noindent\textbf{Tower II: D3-particles.} The ten-dimensional Type IIB string theory allows for the presence of $\text{D}3$-branes.
If these branes are wrapped on some appropriately chosen internal three-cycles, they deliver BPS particles in the external four-dimensional spacetime.

In order to compute the masses of such $\text{D}3$-brane originated particles, let us first recall that the Nambu-Goto part of the effective action describing the dynamics of a generic $\text{D}p$-brane, neglecting the worldvolume gauge fields, is \cite{Ibanez:2012zz}
\begin{equation}
    \label{ExTor_SNG}
    S_{\text{NG},\text{D}p} = \frac{(\alpha')^{-\frac{p+1}{2}}}{(2\pi)^p g_{\text{s}}} \int_{\mathcal{W}} {\rm d} \xi^{p+1} \sqrt{ -h} \,.
\end{equation}
Here $\xi^\iota$, with $\iota = 1, \ldots, p+1$ are the $\text{D}p$-brane worldvolume coordinates, and $h$ is the determinant of the metric $h_{\iota\kappa}$, pull-back of the ten-dimensional spacetime metric \eqref{ExTor_dsans} over the $\text{D}p$-brane worldvolume $\mathcal{W}$.
The prefactor in \eqref{ExTor_SNG} specifies the mass of $\text{D}p$-brane, which is solely dilaton-dependent at the ten-dimensional level.
Particularizing to the case $p=3$, and upon dimensionally reducing \eqref{ExTor_SNG} over the metric ansatz \eqref{ExTor_dsans}, we find that the mass of a $\text{D}3$-brane-originated particle wrapping an internal three-cycle $\Gamma$ is
\begin{equation}
    \label{ExTor_MD3gen}
    M^{\text{D}3}(\Gamma) = \frac{M_{\text{P}}}{4 \pi^{\frac32}} \frac{\text{vol}_3(\Gamma)}{V_Y}\,,
\end{equation}
where $\text{vol}_3(\Gamma)$ denotes the Einstein-frame volume of the wrapped three-cycle $\Gamma$.

Focusing on the regime of small string coupling and large complex structure and K\"ahler moduli, the lightest among the $\text{D}3$-particles that could first break down the effective theory is the one wrapped over the three-cycle Poincar\'e dual to the $\alpha_0$ three-form in \eqref{ExTor_sympl_basis}.
The mass of such lightest $\text{D}3$-particle is
\begin{equation}
    \label{ExTor_MD3light}
    M^{\text{D}3}_\text{lightest} = \frac{M_{\text{P}}}{4 \pi^{\frac32} \sqrt{s^1 s^2 s^3}}\,.
\end{equation}

\noindent\textbf{Tower IIIa: D5-membrane states.} The $\text{D}5$-branes that populate the ten-dimensional Type IIB string theory may deliver BPS domain walls in the external spacetime, and their oscillatory modes constitute an additional, potential source of effective-theory breakdown.
The tension $\mathcal{T}^{\text{D}5}$ of $\text{D}5$-brane originated BPS domain walls can be obtained from \eqref{ExTor_SNG}, after dimensional reduction over the metric ansatz \eqref{ExTor_dsans} and particularizing to the case $p=5$:
\begin{equation}
    \label{ExTor_MD5gen}
    \mathcal{T}^{\text{D}5}(\Gamma) = \frac{M^3_{\text{P}} \ell_{\text{s}}^6}{32 \pi^{\frac72} } \frac{\sqrt{g_{\text{s}}} \; \text{vol}_3(\Gamma)}{\sqrt{s^1 s^2 s^3} V_Y^{\frac32}}\,.
\end{equation}
We shall conventionally take the mass of the oscillatory modes associated to the $\text{D}5$-domain walls as given by $M^{\text{D}5}(\Gamma) = (\mathcal{T}^{\text{D}5}(\Gamma))^{\frac13}$ \cite{Lanza:2020qmt}.
As for $\text{D}3$-particles, in the regime of interest the lightest $\text{D}5$-domain walls are those wrapped on the three-cycle dual to the $\alpha_0$ three-form in \eqref{ExTor_sympl_basis}, and has tension:
\begin{equation}
    \label{ExTor_MD5light}
    \mathcal{T}^{\text{D}5}_\text{lightest} = \frac{M^3_{\text{P}}}{32 \pi^{\frac72} } \frac{\sqrt{g_{\text{s}}}}{v_1 v_2 v_3 \sqrt{s^1 s^2 s^3}}\,,
\end{equation}
with the associated mass scale $M^{\text{D}5}_\text{lightest} = (\mathcal{T}^{\text{D}5}_\text{lightest})^{\frac13}$.

\noindent\textbf{Tower IIIb: NS5-membrane states.} Finally, the ten-dimensional theory Type IIB string theory may be endowed with NS5 brane that, analogously to $D5$-branes, may deliver supersymmetric domain walls once wrapping internal three-cycles.
The tension of ${\text{NS}5}$-originated BPS domain walls can be most readily obtained from the tension of $\text{D}5$-domain walls, after diving it by the string coupling $g_{\text{s}}$:
\begin{equation}
    \label{ExTor_MNS5light}
    \mathcal{T}^{\text{NS}5}_\text{lightest} = \frac{M^3_{\text{P}}}{32 \pi^{\frac72} } \frac{1}{v_1 v_2 v_3 \sqrt{g_{\text{s}} s^1 s^2 s^3}}\,,
\end{equation}
and the associated mass scale is $M^{\text{NS}5}_\text{lightest} = (\mathcal{T}^{\text{NS}5}_\text{lightest})^{\frac13}$.
As such, for small string coupling, we expect the lightest ${\text{NS}5}$-brane states to be always heavier that the lightest ${\text{D}5}$-brane states.

\subsection{Partitioning the moduli space}
\label{sec:Ex_Tor_slice}

With the possible effective theory breakdown sources identified above, let us now show how the moduli space is sliced according to the UV-cutoff.

\begin{figure}[t]
	\centering
	\includegraphics[width=13cm]{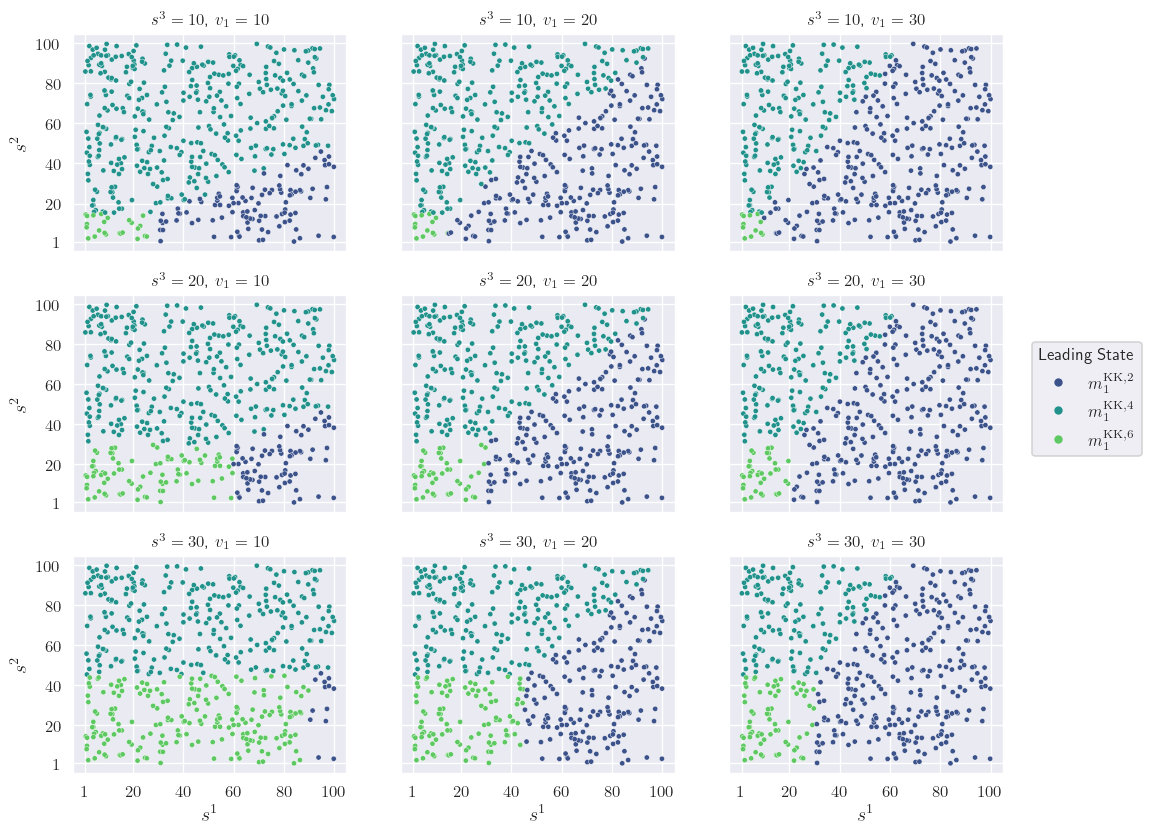}
	\caption{A random set of $500$ points in the two-dimension moduli subspace spanned by the complex structure saxions $s^1$ and $s^2$, in the range $[1,100]$, for the indicated value of the third complex structure saxion $s^3$, and K\"ahler modulus $v_1$. 
	The different shapes and colors of the points label which type of states, among those listed in Section~\ref{sec:Ex_Tor_scales}, first breaks down the effective theory.
	The other moduli, not shown in the plot, have been assumed to have the following, fixed vev's: $v_2 = 20$, $v_3 = 30$, $\phi = -10$.
	\label{Fig:Torus_Data_noObj}}
\end{figure}

\subsubsection{Partition with no fundamental objects}
\label{sec:Ex_Tor_slice_I}

\begin{figure}[!ht]
	\centering
	\includegraphics[width=13cm]{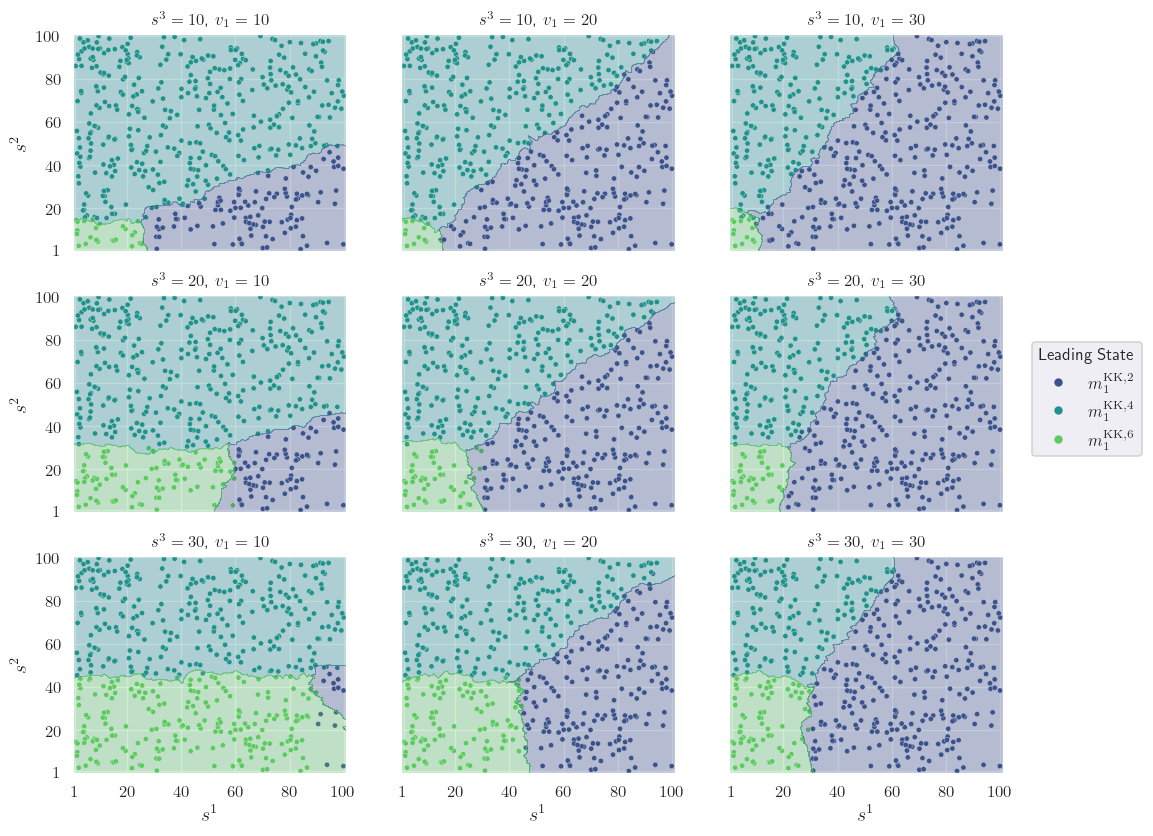}
	\caption{The application of the $k$-nearest neighbor algorithm to the datasets in  Figure~\ref{Fig:Torus_Data_noObj}.
		\label{Fig:Torus_kNN_noObj}}
\end{figure}
\begin{figure}[!ht]
	\centering
	\includegraphics[width=13cm]{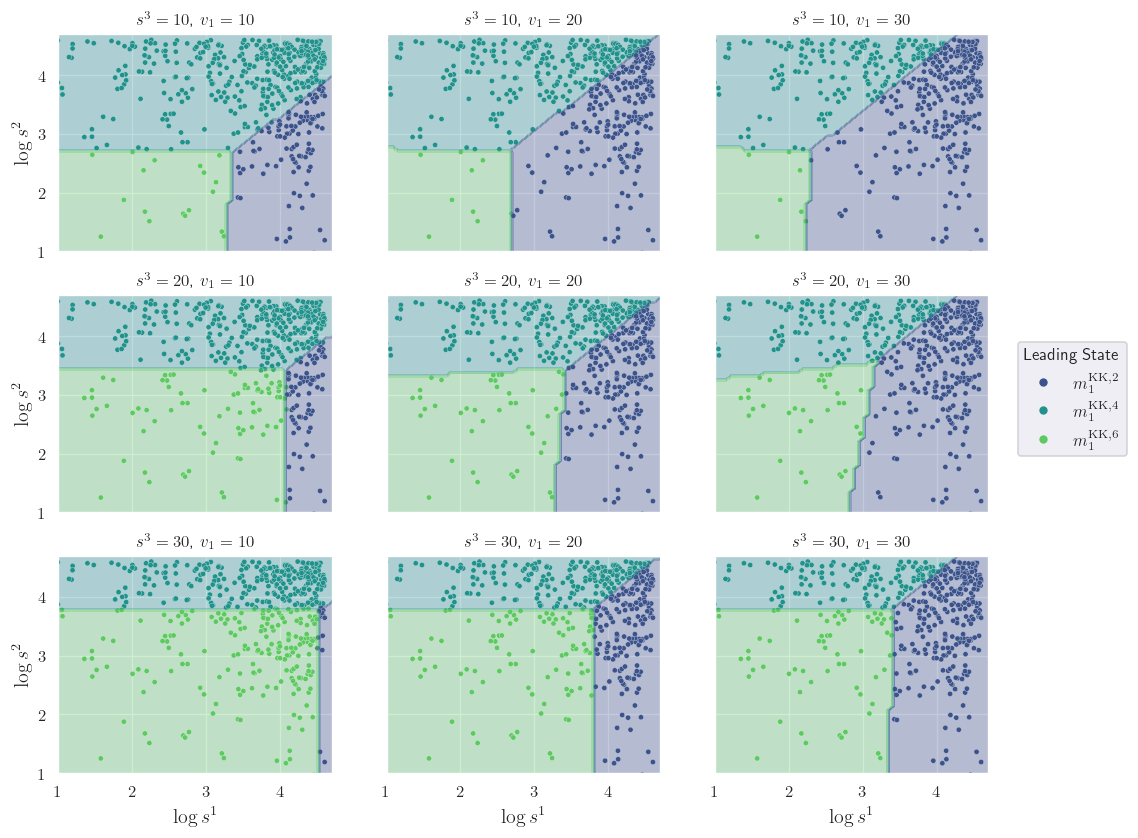}
	\caption{The application of the linear support vector machine algorithm to the datasets in  Figure~\ref{Fig:Torus_Data_noObj}.
		\label{Fig:Torus_LSVM_noObj}}
\end{figure}

To begin with, let us first consider the case where no internal brane is present.
The types of state that can break down the effective theory can then solely be the lightest among the Kaluza-Klein or the winding modes listed in Section~\ref{sec:Ex_Tor_scales}.

For the ease of exposition, and clarity of data visualization, we focus on a two-dimensional subset of the moduli space at a time.
Within each of these subsets we generate a synthetic dataset \eqref{MLDC_data} as follows: we first randomly generate several hundred of points in the chosen subset; then, we will calculate which state, among those listed in Section~\ref{sec:Ex_Tor_scales} is the lightest and would first break down the effective theory.
This procedure can be carried out by setting up a \texttt{python} code, employing the \texttt{pandas} library for handling large datasets.

An example of such a synthetic dataset is portrayed in Figure~\ref{Fig:Torus_Data_noObj}.
Therein, we considered $500$ randomly generated points in the saxionic complex structure subspace spanned by $s^1$ and $s^2$.
Furthermore, we considered three different values for the complex structure modulus $s^3$ and K\"ahler modulus $v_1$, with all the other moduli fixed at given values.
For each of these points we then highlighted, using different colors, which is the type of state that first breaks down the four-dimensional effective field theories among the Kaluza-Klein and winding modes in \eqref{ExTor_mKK_b} and \eqref{ExTor_mw_b}.
Ostensibly, in the regime we considered, the lightest among the Kaluza-Klein modes $m^{\text{KK}, 2i}_k$ -- namely, $m^{\text{KK}, 2i}_1$ --  are the states that break down the effective description first.

Then, we feed the datasets plotted in Figure~\ref{Fig:Torus_Data_noObj} to the $k$-nearest neighbor algorithm and the linear support vector machine algorithm as illustrated in Section~\ref{sec:ML_and_DC_k-nearest} and Section~\ref{sec:ML_and_DC_SVM}, respectively.
In particular, $80\%$ of the data points of Figure~\ref{Fig:Torus_Data_noObj} form the training set for the algorithm, while the remaining ones constitute the test set.
After applying the algorithms, then we can determine how the regions for which each type of states that break down the effective theory look like.

The applications of these algorithms are contained in Figures~\ref{Fig:Torus_kNN_noObj} and~\ref{Fig:Torus_LSVM_noObj}. 
As is clear from the plots, the two algorithms act differently on the datasets of Figure~\ref{Fig:Torus_Data_noObj}: the $k$-nearest neighbor algorithm, whose delivered results are contained in Figure~\ref{Fig:Torus_kNN_noObj}, delivers decision boundaries that look rather irregular.
Instead, the linear support vector machine algorithm, which produces the plots in Figure~\ref{Fig:Torus_LSVM_noObj}, delivers decision boundaries that are just lines, and they can be helpful in guessing, from a bottom-up perspective, the functional form of the masses of the states breaking down the effective theory.

\subsubsection{Partition with fundamental objects}
\label{sec:Ex_Tor_slice_II}

\begin{figure}[t]
	\centering
	\includegraphics[width=13cm]{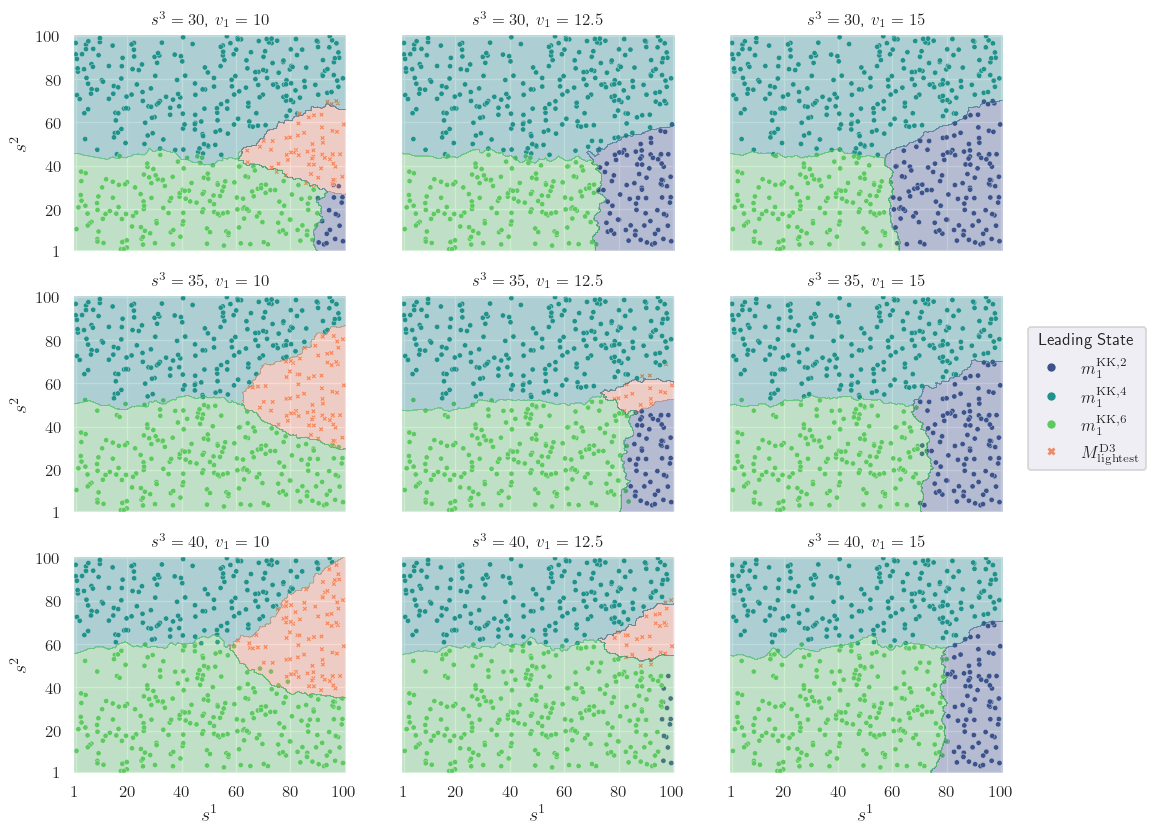}
	\caption{The application of the $k$-nearest neighbor algorithm to a sample of 500 data points, in presence of extended objects.
		\label{Fig:Torus_kNN_withObj}}
\end{figure}
\begin{figure}[!ht]
	\centering
	\includegraphics[width=13cm]{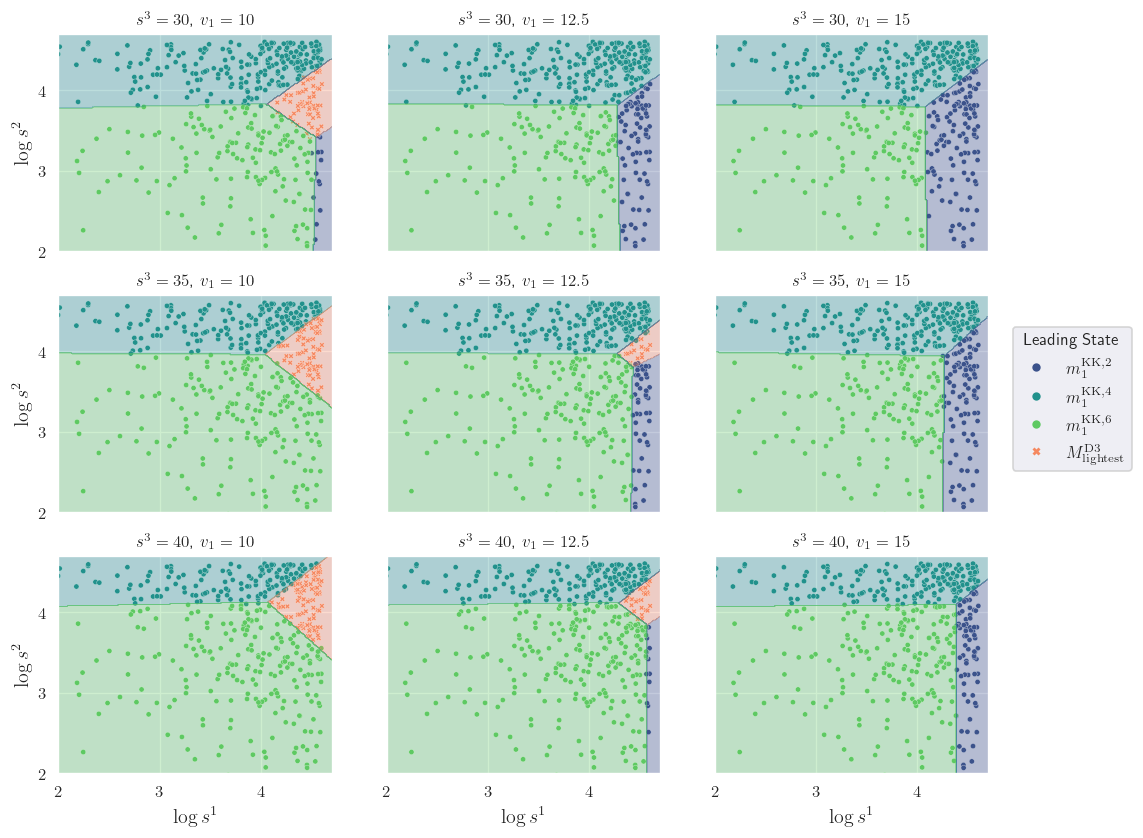}
	\caption{The application of the linear support vector machine algorithm to a sample of 500 data points, in presence of extended objects.
		\label{Fig:Torus_LSVM_withObj}}
\end{figure}

In the presence of fundamental objects, the effective theory might be broken down earlier, at least in some portions of the moduli space, by the states introduced by such objects.
Then, let us assume that the effective four-dimensional theory is populated by particles originated by $\text{D}3$-branes, and by domain walls, stemming from either $\text{D}5$ or $\text{NS}5$-branes, and let us repeat the same procedure performed in the previous section to identify the effective theory cutoff.

In Figure~\ref{Fig:Torus_kNN_withObj} and~\ref{Fig:Torus_LSVM_withObj} are illustrated, respectively, the applications of the $k$-nearest neighbor algorithm and the linear support vector machine algorithm for a data sample of $500$ points in the saxionic space $s^1$, $s^2$, for the indicated values of the complex structure saxion $s^3$ and the K\"ahler modulus $v_1$.
The remaining moduli have been assumed to take the following values: $v_2 = 20$, $v_3 = 30$, $\phi = -1$.

Indeed, at some data points (indicated by crosses, and orange-colored) the state that breaks down first the effective theory is a $\text{D}3$-brane state.
It is worth recalling that $\text{D}3$-brane states may also come in infinite towers (see, for instance, \cite{Grimm:2018cpv,Grimm:2018ohb}), and such a breakdown cannot be solved by including a finite number of states.
For some other regions, instead, the state that breaks down the effective description first is the lightest mode of a Kaluza-Klein tower, and the discussion is no different from the one in the previous Section~\ref{sec:Ex_Tor_slice_I}. 
Interestingly, $\text{D}5/\text{NS}5$-states are never the lightest ones, consistently with \cite{Alvarez-Garcia:2021pxo}.

\section{Conclusions}

In this work we have illustrated how some difficulties in the definition of a cutoff of an effective field theory are ameliorated when the effective theory is tame.
The tame structure of the effective theory allowed us to conclude that only a finite number of different states is required in order to define the effective-theory cutoff throughout its moduli space.
Accordingly, the moduli space is partitioned into a finite number of subsets, in each of which a different state is the relevant one that determines the cutoff.
Furthermore, we have shown that this partition can be concretely determined by employing supervised machine learning techniques, only by knowing how the effective theory is broken down at some points in the moduli space.

The ideas presented in this work can be expanded, or tested following several directions.
To begin with, the support vector machine algorithm that we have utilized throughout this work is a linear one.
However, as explained in Section~\ref{sec:ML_and_DC_SVM}, such a linear method is, in particular, useful in two circumstances: over the full patch of the moduli space whenever the states determining the cutoff are strictly monomials, or towards the boundaries of the moduli space if the couplings are monomially tamed.
If the couplings are more complicated functions of the moduli -- for instance, they could be polynomially tamed -- different support vector machine algorithms should be employed in order to obtain a meaningful partition of the moduli space.

It would be interesting to additionally test the construction of the moduli space partition proposed here to effective field theories more complicated than the toroidal model considered in Section~\ref{sec:Ex_Tor}.
In particular, it would be nice to get a partition when the masses of the states determining the effective-theory cutoff exhibit a more involved dependence on the axion fields populating the theory, and check to what extent they participate in determining the partition.
It should be stressed, however, that constructing a synthetic set of data points as performed in Section~\ref{sec:Ex_Tor} for the toroidal model requires a clean control over the numerical factors that appear in the masses of the effective-theory breakdown states, which for some models may be hard to achieve.

\vspace{1em}
\noindent{\textcolor{colorloc5}{\textbf{Acknowledgments}}}

\noindent I am deeply grateful to Thomas Grimm for continuous support and suggestions throughout the writing of this work, and to Mick van Vliet and Timo Weigand for precious comments on the draft. I would also like to thank Florent Baume, Cesar Fierro Cota and Jeroen Monnee for interesting discussions.

\noindent This research is supported in part by Deutsche Forschungsgemeinschaft under Germany’s Excellence Strategy EXC 2121 Quantum Universe 390833306 and by Deutsche Forschungsgemeinschaft through a German-Israeli Project Cooperation (DIP) grant “Holography and the Swampland”.  During the initial stage of the project, this research was partly supported by the Dutch Research Council (NWO) via a Start-Up grant and a Vici grant.

\appendix

\section{An overview of the \texorpdfstring{$k$}{k}-nearest neighbor algorithm}
\label{sec:k-near_review}

In this section we overview the $k$-nearest neighbor algorithm that we have introduced in Section~\ref{sec:ML_and_DC_k-nearest} and then extensively employed in Section~\ref{sec:Ex_Tor} for examining the toroidal orbifold case.
For a string theory-oriented, more complete review of the algorithm and related ones we refer to \cite{Ruehle:2020jrk}.

Assume that we have some data that is labeled by some variables (`\emph{features}'), that we collect in the vector ${\bf x}^{(i)}$, with the index $i$ labeling the data entry.
We also assume that the data are organized in classes (the `\emph{labels}'), that we label as $y^{(i)}$. 
The values that $y^{(i)}$ may take belong to a set $C$ with finite cardinality.
Conventionally, we may take $C = \{ 0, 1, \ldots, N \}$, with $N$ being the set cardinality.
Thus, the dataset we are focusing on can be represented by the set
\begin{equation}
	\label{kNN_data}
	\text{Data} = \{ ({\bf x}^{(i)},\, y^{(i)}) \}\,.
\end{equation}

For concreteness, consider the data represented in Figure~\ref{Fig:Example_Data}: the features of the data are the values of the moduli, with ${\bf x}^{(i)}$ being the two-dimensional vector $((s^1)^{(i)},(s^2)^{(i)})$, with $i$ labeling the point in the dataset therein represented.
The target, or the class $y^{(i)}$ is given by the type of the lightest state that emerge at the given moduli space point $((s^1)^{(i)},(s^2)^{(i)})$, and is represented by the color of the points.
More generically, given the dataset \eqref{MLDC_data}, the features ${\bf x}^{(i)}$ correspond to $\varphi^A_{i}$, and the target $y^{(i)}$ corresponds to the $\text{Type}\, J_{i}$.

Now, assume to take a new point in the features space, ${\bf x}_{\text{new}}$. The question that we wish to address is: \emph{based on the knowledge we have about the dataset \eqref{kNN_data}, can we infer what is the most likely class $y_{\text{new}}$ (the `\emph{target}') to which ${\bf x}_{\text{new}}$ belongs?}
This problem can be addressed by means of some supervised machine learning classification algorithms, that can \emph{learn} from the original dataset \eqref{kNN_data} which common properties the features associated to the same class share. 
As stated in Section~\ref{sec:ML_and_DC_k-nearest}, the techniques employed to face the problem at hands are `\emph{supervised}', since in the original dataset \eqref{kNN_data} the points are labeled, namely we already know to which class each of the points ${\bf x}^{(i)}$ belongs to.\footnote{Indeed, classification algorithms may also be \emph{unsupervised}.  However, in these cases we do not have any knowledge the labels $y^{(i)}$, but rather the question addressed is: \emph{given some unlabeled points ${\bf x}^{(i)}$, can we organize them in some classes?} In other words, in this case, the classes (and their number) are not given, but created by the algorithm.}

The $k$-nearest neighbor algorithm is one of the most popular and simplest classification algorithm, and proceeds as follows.
Given the new feature ${\bf x}_{\text{new}}$, we compute the Euclidean distance between ${\bf x}_{\text{new}}$ and each of the features ${\bf x}^{(i)}$ entering the dataset \eqref{kNN_data}. 
Then, we select the $k$ points ${\bf x}^{(i)}_{\text{nearest}}$ that have the shortest distance with respect to ${\bf x}_{\text{new}}$. 
Finally, we check what is the most frequent label $y^{(i)}$ among the points ${\bf x}^{(i)}_{\text{nearest}}$: this most frequent class is the predicted class $y_{\text{new}}$ for the point ${\bf x}_{\text{new}}$.
The decision regions -- such as the one plotted in Figure~\ref{Fig:Example_Partition} -- can be obtained by introducing a (finite) lattice of points in the variable space, and applying the $k$-nearest neighbor to each of the points in the lattice.

Some comments are in order regarding the $k$-nearest neighbor algorithm.
Firstly, as should be clear from how the algorithm works, the algorithm does not operate on the original dataset \eqref{kNN_data}, and does not produce any model out of it alone, unlike several other machine learning techniques. 
Rather, the algorithm operates only when the dataset is queried, namely when we ask what is the class of new points. 
For this reason, the $k$-nearest neighbor algorithm is among the so-called `\emph{lazy}' machine learning algorithms.

Secondly, the value of $k$ is arbitrary, and it has to be carefully chosen. 
A value of $k$ that is too large may render the algorithm too rough: in the limit in which $k$ equals the number of data points, for every new point we would always predict the same class, that is the most frequent in \eqref{kNN_data}.
On the other hand, if the value of $k$ is too small, the algorithm may not be accurate: if $k = 1$, every new data is associated just to the class of the closest point in \eqref{kNN_data}.
Therefore, typically some intermediate, model-dependent choice ought to be taken, and
one could also test different values of $k$ and check which one delivers the most accurate boundary regions.

\section{An overview of the Linear Support Vector Machine algorithm}
\label{sec:LSMVMreview}

Support vector machine algorithms, whose first inception can be traced back to the works of Vladimir Vapnik and collaborators \cite{boser1992support,cortes1995support}, are supervised machine learning algorithms that employ regression techniques to classification problems.

The dataset that is fed to a support vector machine algorithm is of the same type as the one employed for the $k$-nearest neighbor algorithm in \eqref{kNN_data}: namely, we consider a set of data, with each of the entries specified by some feature, or variables ${\bf x}^{(i)}$, and a label, or class $y^{(i)}$.
The question that support vector machine algorithms aim to address is also equivalent to the $k$-nearest neighbor algorithm in \eqref{kNN_data}: \emph{given a new point, ${\bf x}_\text{new}$, what is the class $y_\text{new}$ to which it belongs?}
However, support vector machine algorithms address this question in a substantially different way than the $k$-nearest neighbor algorithm.

For ease of exposition, we will focus on the case where the categorical variable $y$ can take just two values, $y_1$ and $y_2$, and we will conventionally take $y_1 = 1$ and $y_2 = -1$. 
The goal of the linear support vector machine algorithm is to identify the hyperplane, in the space of the features ${\bf x}$ that best separates the data ${\bf x}^{(i)}$ characterized by $y^{(i)} = 1$ from those that belong to the other category, distinguished by $y^{(i)} = -1$.
Let us denote this hyperplane that we wish to find with $H_0$, which can be described by the equation
\begin{equation}
	\label{LSVM_H0}
	H_0: \qquad {\bf w}^T {\bf x} + b = 0\,.
\end{equation}
Here ${\bf w}$ is a vector with the same dimension as the features ${\bf x}$, and $b$ a real number, both to be determined via the algorithm.
We shall assume that the two classes of data are \emph{linearly separable} and, thus, with an hyperplane we can fully separate the two classes, with no point being misclassified.

Now, we introduce two hyperplanes, parallel to $H_0$:
\begin{equation}
	\label{LSVM_H1H2}
	H_1: \qquad {\bf w}^T {\bf x} + b = 1\,, \qquad \qquad H_2: \qquad {\bf w}^T {\bf x} + b = - 1\,.
\end{equation}
The hyperplane $H_0$ we look for lies exactly in the middle of the strip between $H_1$ and $H_2$, and the region between the hyperplane $H_1$ or $H_2$ and $H_0$ is called `\emph{margin}'.
Clearly, knowing the equations of the hyperplanes $H_1$ and $H_2$ \eqref{LSVM_H1H2} allow us to most readily obtain the equation of the hyperplane $H_0$ in \eqref{LSVM_H0}.

The algorithm then proceeds in determining the equations of the hyperplanes $H_1$ and $H_2$ as follows.
The hyperplane $H_1$ is defined in such a way that any point that lies on $H_1$, or above it falls in the category $y_1 = 1$; namely:
\begin{equation}
	\label{LSVM_H1data}
	{\bf w}^T {\bf x}^{(i)} + b \geq 1 \qquad \text{for any ${\bf x}^{(i)}$ with $y^{(i)} = 1$}\,.
\end{equation}
Conversely, all the points in the class $y_2 = -1$ lie on, or below the hyperplane $H_2$:
\begin{equation}
	\label{LSVM_H2data}
	{\bf w}^T {\bf x}^{(i)} + b \leq - 1 \qquad \text{for any ${\bf x}^{(i)}$ with $y^{(i)} = -1$}\,.
\end{equation}
\begin{figure}[th]
	\centering
	\includegraphics[width=9cm]{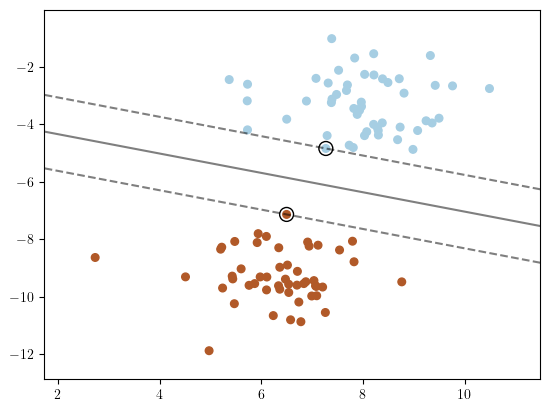}
	\caption{An example of application of the linear support vector machine algorithm. 
		The dataset, composed by $100$ points, is obtained via the \texttt{make\_blobs} library.
		The dashed lines denote the hyperplanes $H_1$ and $H_2$, passing through the two support vectors here encircled.
		The solid line represents the decision boundary $H_0$.
		\label{Fig:App_Example_SVM}}
\end{figure}

In order for the algorithm to be effective and solid in predicting the class of new entries, one should look for the hyperplanes $H_0$ that is located at the largest distance from any point of the classes;
in turn, this implies that the hyperplanes $H_1$ and $H_2$ should be at their largest distance from one another.
The distance between the hyperplanes $H_1$ and $H_2$ -- namely, the `width' of the margin -- is $2/\| {\bf w} \|$; therefore, in order to find the hyperplanes $H_1$ and $H_2$ that deliver the widest margin, we ought to solve the following optimization problem:
\begin{equation}
	\label{LSVM_opt_prob}
	\min\limits_{{\bf w}, b} \| {\bf w} \| \qquad\qquad \text{such that} \quad y^{(i)}({\bf w}^T {\bf x}^{(i)} + b) \geq 1  \qquad \forall\, i\,,
\end{equation}
with the latter condition concisely encoding both \eqref{LSVM_H1data} and \eqref{LSVM_H2data}.
Solving the optimization problem \eqref{LSVM_opt_prob} delivers the parameters ${\bf w}$ and $b$, whence the equation of the decision boundary $H_0$ in \eqref{LSVM_H0} is known.

In concrete machine learning application, the original dataset can be split into a training set and a test set, with, typically, $80\%$ of the data assigned to the former, and the remaining $20\%$ to the latter.
Then, one can apply the optimization procedure formulated in \eqref{LSVM_opt_prob} to the training set, which may be realized with gradient descent methods.
After obtaining the equation for the decision boundary $H_0$, one could test whether such a decision boundary correctly predicts the classes of the data contained in the test set.

A simple example of application of linear support vector machine algorithm is depicted in Figure~\ref{Fig:App_Example_SVM}.
The dataset employed therein has been synthetically obtained employing the \texttt{make\_blobs} library, that creates Gaussian-distributed data belonging to an arbitrary number of classes.
The support vector machine algorithm determines the two dashed lines there depicted via the optimization procedure \eqref{LSVM_H2data}, and they serve as the hyperplanes (here reduced to lines) $H_1$ and $H_2$ determining the margin.
The lines $H_1$ and $H_2$ pass through the support vectors, encircled in Figure~\ref{Fig:App_Example_SVM}, which are the points of each class that are closest to the other class.
Once the lines $H_1$ and $H_2$ are known, the decision boundary $H_0$, the solid line in Figure~\ref{Fig:App_Example_SVM}, is determined.

It is worth mentioning that the linear support vector machine algorithm just explained is one of the simplest incarnation of the algorithm, and several, more sophisticated version thereof have been introduced in the last decades.
For instance, one could look for `\emph{soft margins}' -- rather than the `\emph{hard margins}' as we did above -- which allow for some data to be misclassified, and enter the `wrong' decision region.
This can helpful whenever the problem is not linearly separable, or if we know that some data are anomalous.
Alternatively, one can generalize the algorithm in such a way that more complicated decision boundaries can be delivered, whose equations are not simple hyperplanes. 
The exploration of the latter possibility is left for future work.


\providecommand{\href}[2]{#2}\begingroup\raggedright\endgroup

\end{document}